\documentclass[aps,pra,twocolumn,floatfix,longbibliography,nofootinbib,tbtags]{revtex4-1}
\usepackage[utf8]{inputenc}
\usepackage{natbib}
\usepackage{graphicx}
\usepackage{xcolor}
\usepackage{mathtools}
\usepackage{amsmath}
\usepackage[colorlinks, linkcolor=blue]{hyperref}
\hypersetup{colorlinks,allcolors=black}
\usepackage{amssymb}
\usepackage{gensymb}
\usepackage{float}

\usepackage{tabularx,graphicx}
\usepackage{epstopdf}
\usepackage{latexsym}
\usepackage{color, colortbl}
\usepackage{psfrag}
\usepackage{bbm}
\usepackage{bm}
\usepackage{titlesec}
\usepackage{dsfont}
\usepackage{feynmp}
\usepackage{slashed}
\usepackage{multirow}
\usepackage[normalem]{ulem}
\renewcommand{\vec}[1]{\boldsymbol{#1}}

\def \k {{\vec k}}

\def \v {{\bf v}}

\def \beq {\begin{eqnarray}}
\def \eeq {\end{eqnarray}}

\begin{document}
\title{High temperature transport in the one dimensional mass-imbalanced Fermi-Hubbard model}
\author{Thomas G. Kiely}
\author{Erich J. Mueller}
\affiliation{Department of Physics, Cornell University, Ithaca, New York 14853, USA.}

\begin{abstract}
We study transport in the one-dimensional mass-imbalanced Fermi-Hubbard model 
% at infinite temperature
{%\color{blue}
in the high-temperature limit}, focusing on the case of strong interactions. Prior theoretical and experimental investigations have revealed unconventionally long transport timescales, with complications due to strong finite size effects. 
%Using novel tensor network techniques, we compute the current-current correlation function directly in the thermodynamic limit. By analytically expanding about the limit of strong interactions and imbalances we clarify the role of long-lived oscillations in transport properties. We additionally propose an experimental sequence to measure these correlation functions in cold atom experiments.
We compute the dynamical current-current correlation function directly in the thermodynamic limit using infinite tensor network techniques. We show that transport in the strong-imbalance limit is dominated by AC resonances, which we compute with an analytic expansion. We study the dephasing of these resonances with mass imbalance, $\eta$. In the small-imbalance limit, the model is nearly integrable. We 
%demonstrate the interpolation between 
connect
these unusual limits by computing the DC conductivity and transport decay time as a function of $\eta$ and the interaction strength $U/t$. We %additionally 
propose an experimental protocol
%sequence 
to measure these correlation functions in cold atom experiments.
\end{abstract}

\maketitle

\section{Introduction}
The one-dimensional (1D) mass-imbalanced Fermi-Hubbard model, where $\uparrow$-spin particles hop more easily than $\downarrow$-spins, interpolates between two interesting limits. When the masses are equal the system is integrable. When the mass ratio diverges the heavy particles act as stationary disorder, localizing the light particles. In neither of these limits does the system have typical metallic behavior. Between %these limits, 
them
the system is non-integrable, and should display conventional metallic behavior,
but it is challenging to calculate the conductivity when the interactions are strong.
%For strong interactions it is challenging to calculate the conductivity and the relaxation time. 
In this paper we use infinite tensor network methods and analytic expansions to study the high temperature optical conductivity of this model.

In a conventional (non-integrable) metallic system, the zero-temperature charge conductivity exhibits a zero-frequency delta-function peak whose area is the Drude weight~\cite{scalapino1993}. At any temperature $T>0$, scattering processes broaden this peak, and the Drude weight vanishes.  Instead, the conductivity at frequency $\omega$ has the approximate Drude form $\sigma(\omega)=\sigma_{\rm DC}/(1-i\omega\tau)$ with a transport scattering time $\tau$, and DC conductivity $\sigma_{\rm DC}$. Integrable systems are characterized by an infinite number of conserved quantities, and consequently violate this simple picture~\cite{bertini2021}. For example, the ordinary (mass-balanced) 1D Fermi-Hubbard model exhibits a finite Drude weight at {\it all} temperatures so long as the total charge density, $\bar n$, is not  unity~\cite{ilievski2017}. This feature corresponds to an infinite DC conductivity, which in higher dimensions is only seen in superfluids or zero-temperature metals. At half-filling $(\bar n=1)$, the Drude weight vanishes but the high temperature transport is still unconventional, displaying a {%\color{blue}
Kardar-Parisi-Zhang} (KPZ) dynamical scaling~\cite{znidaric2011,denardis2019,ljubotina2019,gopalakrishnan2019,wei2022} which has recently been attributed to a non-Abelian ${\rm SO}(4)$ symmetry~\cite{fava2020}.

The limit of strong mass imbalance also displays unusual transport. When the heavy-particle hopping vanishes, local heavy-particle densities are constants of motion, and hence the system is integrable. This integrable limit has vastly different properties than the symmetric-mass limit.   
The static heavy particles act as disorder, leading to Anderson localization of the light particles \cite{anderson1958}.  In one dimension this localization occurs for any non-zero interaction strength.
%In one dimension, the presence of this ``background disorder" (i.e. a thermal ensemble-average over heavy-particle positions) Anderson-localizes the light-particle eigenstates. 
In a compelling but incorrect argument, it was proposed that analogous physics might be found if the heavy particles were allowed to hop, leading to
%this thermally-activated disorder might localize the light particles at a finite mass imbalance, which could be understood as 
many-body localization in a translationally-invariant system~\cite{deroeck2014,Grover_2014,schiulaz2015}. Subsequent work, however, provided evidence that the model is ergodic for any finite mass ratio and interaction strength~\cite{deroeck2014b,PAPIC2015714,yao2016,sirker2019}. Nonetheless, the system displays ``anomalously long" decay times~\cite{jin2015,zechmann2022}, and there is a time-scale over which the behavior appears non-diffusive. One essential feature is that the long-time limit ($\tau\to\infty$) does not commute with the thermodynamic limit ($L\to\infty$), and results from finite-size numerics are only reliable for short times. In our numerical calculations, we leverage tensor network techniques to study transport properties directly in the thermodynamic limit, circumventing this challenge.  

Ultracold atom experiments have recently realized the mass-imbalanced Fermi Hubbard model~\cite{darkwah2022}.  Fermionic Ytterbium atoms are trapped in a 2D near-resonant optical lattice.  Due to the spin-dependent AC polarizability, atoms with different internal states see lattices of different depths, and hence have different effective masses~\cite{riegger2018}.  The mass ratio depends on the frequency of the lattice lasers and approaches unity for a far-detuned lattice.
{%\color{blue} 
Such state-dependent lattices have also been realized using 
fermionic
Strontium~\cite{heinz2020} and Potassium~\cite{jotzu2015} atoms as well as bosonic Rubidium~\cite{McKay_2010}.}
%can be reduced by moving the lattice frequency farther from resonance.  
Interactions are tuned by a Feshbach resonance~\cite{bloch2008,chin2010}. In these experiments, conventional transport observables (e.g. DC resistivity) are not easily accessible. Instead, transport is  studied by introducing spin or charge deformations and observing their relaxation~\cite{demarco2019,thywissen2019,Brown2019,darkwah2022,gross2017}.  We propose, and model, a novel technique to extract the current-current correlation function from the dynamical response to a probe. 

In Section~\ref{sec:formalism}, we introduce the mass-imbalanced Fermi-Hubbard model and discuss the transport properties studied here. In Sec.~\ref{sec:results}, we present our numerical results. In Sec.~\ref{sec:analytic} we perform an analytic expansion about the large-mass-ratio and strong-interaction limit, which is shown to quantitatively model the transport properties. In Sec.~\ref{sec:experiment} we discuss experimental implications of our work, including a proposal for measuring the current-current correlation function in ultracold atomic systems. Our conclusions are presented in Sec.~\ref{sec:discussion}.

\section{Formalism\label{sec:formalism}}

\subsection{Fermi-Hubbard model}
The mass-imbalanced Fermi-Hubbard model is defined by the Hamiltonian
\begin{multline}
    H=-\sum_{i\mu} \left( t_\mu c^\dagger_{i,\mu}c_{i+1,\mu}+{\rm H.c.}\right)    +U\sum_in_{i,\uparrow}n_{i,\downarrow}
    \label{eq:imbFH}
\end{multline}
where $i$ labels the sites, $\mu=\uparrow,\downarrow$ labels the spins, and
$t_\mu$ and $U$ parameterize the kinetic and interaction energies, respectively.  We will also write $t_\uparrow=t$ and $t_\downarrow=\eta t$, so that $\eta$ parameterizes the mass imbalance. We take $0\leq\eta\leq 1$,  defining $\downarrow$ spins as the heavy particles.

As already explained, in the absence of any mass imbalance ($\eta=1$), the model is integrable and can be solved exactly with the Bethe ansatz~\cite{lieb1968,schultz1990}. %For total filling fractions $\bar n\neq 1$, the mass-balanced Fermi-Hubbard model has a finite Drude weight (i.e. an infinite DC conductivity), even at infinite temperature~\cite{bertini2021}. The half-filled ($\bar n=1$) case is a high-symmetry point and has been argued to have zero Drude weight and unconventional superdiffusive behavior~\cite{bertini2021,fava2020}.
Adding a mass imbalance ($\eta<1$) formally breaks integrability.
%, and one expects that the Drude weight vanishes at finite temperature for all fillings.

In the limit of infinite mass imbalance ($\eta=0$), Eq.~(\ref{eq:imbFH}) 
reduces to the Falicov-Kimball model~\cite{falicov1969}, which has an extensive number of local conserved densities: $[H,n_{i,\downarrow}]=0~\forall~i$.  One can think of the model as describing non-interacting spinless fermions interacting with a static binary potential given by the configuration of heavy spins. At high temperatures the thermal density matrix sums over all possible binary disorder configurations, and the model is expected to exhibit Anderson localization for $\eta=0$~\cite{jin2015,heitmann2020}. For finite $\eta\to 0$, the model is ergodic but with a diverging relaxation time~\cite{deroeck2014b,PAPIC2015714,yao2016,sirker2019,zechmann2022}.

We report on the case of {\em half filling} where the ensemble average number of particles on a site are {%\color{blue} 
$\langle n_{i\mu}\rangle=1/2$.  
% Results at other densities are similar.
In the $\eta=1$ model, the half-filled system has an enhanced symmetry with respect to other fillings, resulting in unconventional subdiffusive dynamical scaling~\cite{fava2020}. This makes the half-filled case particularly interesting. At other fillings, transport is ballistic when $\eta=1$~\cite{ilievski2017}. For generic $\eta\neq 1$, we expect that results at different densities are qualitatively similar.}

\subsection{Transport}

In this paper we quantify transport by studying the behavior of the optical conductivity. Using the fluctuation-dissipation theorem~\cite{kubo}, one can express the real part of the spin- and site-resolved conductivity as
\begin{equation}
    \sigma_{\mu\nu}(l,m;\omega)=\frac{1-e^{-\beta\omega}}{2\omega}\int_{-\infty}^\infty d\tau~e^{i\omega\tau}\Lambda_{\mu\nu}(l,m;\tau),
    \label{eq:kubo_site}
\end{equation}
where $\Lambda_{\mu\nu}(i,j;\tau)$ is the current-current correlation function:
\begin{equation}
    \Lambda_{\mu\nu}(l,m;\tau)={\rm Tr}\big(e^{-\beta H}j_\mu(l,\tau)j_\nu(m,0)\big).
\end{equation}
The current operator acting on sites $(l,l+1)$ is defined as
\begin{equation}\label{eq:j_local}
    j_\mu(l)=-it_\mu\left(c^\dagger_{l+1,\mu}c_{l,\mu}-{\rm H.c.}\right),
\end{equation}
and $j_\mu(l,\tau)=e^{iH\tau}j_\mu(l)e^{-iH\tau}$. Of course, the conventional expression for the optical conductivity is recovered by summing over indices $l$ and $m$:
\begin{equation}
    \sigma(\omega)=\frac{1}{N}\sum_{l,m}\sigma(l,m;\omega)
\end{equation}

We will mainly be concerned with $\sigma=\sigma_{\uparrow\uparrow}$,
and for notational simplicity will denote $j=j_\uparrow$, and $\Lambda=\Lambda_{\uparrow\uparrow}$, omitting the subscripts when we are referring to the light particles. We make this choice because the high-temperature charge ($\sigma_c$) and spin ($\sigma_s$) conductivities are both directly proportional to $\sigma_{\uparrow\uparrow}$ when $\eta=0$ and $\eta=1$.

%One can in principle also study the other components of the conductivity tensor, $\sigma_{\downarrow\downarrow}$ and $\sigma_{\uparrow\downarrow}$. For simplicity, we focus on $\sigma_{\uparrow\uparrow}$ because it reduces to the total conductivity at the points $\eta=0$ and $\eta=1$ (up to a prefactor).

%  This is the natural quantity to consider when $\eta$ is small.

% As we are primarily concerned with the $\eta\to 0$ limit, we will take $J=J_\uparrow$ to be the current operator for the lighter spin-species,
% \begin{equation}
%     J_\uparrow=-it\sum_{j}\left(c^\dagger_{j+1,\uparrow}c_{j,\uparrow}-{\rm H.c.}\right),
% \end{equation}
% which of course reduces to the total charge current operator as $\eta\to 0$. We will henceforth drop the subscript $\uparrow$ from the current operator. 

For systems with a bounded spectrum, as we consider here, we can expand Eq.~(\ref{eq:kubo_site}) in the high-temperature limit, $T\gg t,U$~\cite{lindner2010,pereplitsky2016,kiely2021,patel2022}:
\begin{equation}
    \sigma(l,m;\omega)=\frac{\beta}{2}\int_{-\infty}^\infty d\tau~e^{i\omega \tau}~{\rm Tr}\big(j(l,\tau)j(m,0)\big)+\cdots.
    \label{eq:highTSigma}
\end{equation}
Thus, up to an overall factor of $\beta$, the high-temperature optical conductivity is simply the Fourier transform of the current-current correlation function, $\Lambda(l,m;\tau)={\rm Tr}\big(j(l,\tau)j(m,0)\big)$. This expansion can be performed for $\sigma_{\uparrow\uparrow}$ and $\sigma_{\downarrow\downarrow}$, but the cross terms, $\sigma_{\uparrow\downarrow}$, are only non-zero at second order ($\propto\beta^2$). Here we focus on the leading-order expansion, studying $\Lambda(\tau)$.

% {\color{blue} In real physical systems, the 1D single-band Fermi-Hubbard Hamiltonian emerges as a low-energy description of lattice fermions. Hence, the physics we capture in this high-temperature expansion takes place at temperature $T$ that are large compared to the bandwidth but small compared to the band-gap. In previous work~\cite{kiely2021}, the present authors have argued that such a regime is relevant for modern experiments on ultracold fermions in an optical lattice.}

Going forward, we will present calculations performed on uniform, infinite-temperature systems in the thermodynamic limit. Making use of translational invariance, we will define $\Lambda(x,\tau)=(1/N)\sum_{l}\Lambda(l,l+x;\tau)$, which is well-defined in the limit $N\to\infty$. In one dimension, the units of the current-current correlation function are those of a squared characteristic rate, and hence we will report it in units of $t^2$. We will report values for the high-temperature conductivity in units of $\sigma_{\rm ref}=a\beta t$ where $a$ is the lattice spacing. This corresponds to the conductivity from a random diffusive walk with scattering time $1/t$ and a mean free path $a$. 

%In solid-state systems, the spatial (momentum) dependence of $\Lambda$ is often neglected because the measured frequency $\omega$ is much less than the Fermi energy. In these systems, it often suffices to take the uniform $k\to 0$ limit. In cold atom quantum simulators, by contrast, one can measure correlations on timescales $\sim1/t$, and hence the momentum dependence of the conductivity may become relevant. In this work, we compute $\Lambda(x,\tau)$ directly in the thermodynamic limit, as described in the next section. 
%While we will focus most of our attention on the uniform limit, we present a complete picture of the correlation functions 

\subsection{Tensor Network Techniques\label{sec:numericalTechniques}}
We compute the real-time, spatially-resolved current-current correlation function $\Lambda(x,\tau)$ at infinite temperature using infinite tensor network techniques \cite{SCHOLLWOCKreview,PAECKEL2019}. 
%Much of our analysis can be found in the standard toolbox of matrix product state manipulations, and we refer readers to Refs.~\cite{SCHOLLWOCKreview,PAECKEL2019} for a more in-depth discussion.
%The Hilbert space of each site of the optical lattice is four dimensional, spanned by $|0\rangle,c_\uparrow^\dagger |0\rangle,c_\downarrow^\dagger |0\rangle,c_\uparrow^\dagger c_\downarrow^\dagger |0\rangle$.  We will write these four basis states as $|s\rangle$ with $s=1,2,3,4$.
%In the infinite-temperature limit, the thermal density matrix $\rho_{\beta}=e^{-\beta H}$ approaches the identity, $\rho_0=\mathbb{I}$, which can be expressed as
%a matrix product operator
%. This implies that all fluctuations are uncorrelated, allowing one to factor the density matrix into an outer product of site-local terms. In other words, if an index $s_i$ denotes the states of the local Hilbert space on lattice site $i$, then the infinite-temperature density matrix can be written as
%\begin{equation}
%     \rho_0=\sum_{\vec s,\vec s^\prime}\left(\prod_i\otimes~ \delta_{s_i,s^\prime_i}~|s_i\rangle \langle s_i^\prime| \right)
%     \label{eq:rho0}
% \end{equation}
% where the sum over $\vec s$ denotes a sum over all possible values of $s_j$.  We can purify this operator, writing it as
%
%$\{s_1,s_2,s_3\ldots\}$ and $\delta_{i,j}$ is the Kronecker delta function.
%
As described in Appendix~\ref{sec:timeEvol}, we purify the infinite-temperature density matrix, writing it as $\rho_0=\rm Tr|\psi_0\rangle\langle \psi_0|$ where the trace is taken over a set of auxiliary degrees of freedom, corresponding to a copy of the original system~\cite{SCHOLLWOCKreview,PAECKEL2019}. We write $|\psi_0\rangle$ as an infinite matrix product state.

Operator expectation values can be written as ${\rm Tr}(\rho_0\hat O)=\langle\psi_0|\hat O_{\rm phys}|\psi_0\rangle$ where $\hat O_{\rm phys}$ acts only on the physical degrees of freedom of $|\psi_0\rangle$. Given this definition, we may write the current-current correlation function as an expectation value:
\begin{equation}
    \Lambda(x,\tau)=\langle \psi_0|e^{iH_{\rm phys}\tau}j_{\rm phys}(0) e^{-iH_{\rm phys}\tau}j_{\rm phys}(x)|\psi_0\rangle
    \label{eq:lambda1}
\end{equation}
where $j_{\rm phys}(0)$ is a local current operator of the form in Eq.~(\ref{eq:j_local}) connecting the {\it physical} sites at $x=0$ and $x=1$. As the infinite-temperature density matrix is simply the identity operator, its purification has a special property: for any operator $\hat O_{\rm phys}$ acting on the physical degrees of freedom, there is a unique operator $\hat{O}^\prime_{\rm aux}$ acting only on the auxiliary degrees of freedom which satisfies $\hat O_{\rm phys}|\psi_0\rangle=\hat O_{\rm aux}^\prime|\psi_0\rangle$~\cite{KENNES2016}. The relationship between $\hat O$ and $\hat O^\prime$ is determined by the choice of purification, $|\psi_0\rangle$, which is not unique. In Appendix~\ref{sec:timeEvol} we elaborate on how the choice of purification can be useful in implementing block-sparsity constraints~\cite{kiely2022}, and we 
%give a systematic procedure for determining 
show how to determine the auxiliary partner $\hat O^\prime$ to an operator $\hat O$ for a given purification.

Given this property of $|\psi_0\rangle$, we define $H^\prime_{\rm aux}$ as the auxiliary partner of the Hamiltonian, $H_{\rm phys}$. This allows us to rewrite Eq.~(\ref{eq:lambda1}) as
\begin{equation}
    \Lambda(x,\tau)=\langle \psi_0|j_{\rm phys}(0) e^{-i(H_{\rm phys}-H^\prime_{\rm aux})\tau}j_{\rm phys}(x)|\psi_0\rangle,
    \label{eq:lambda2}
\end{equation}
where we have taken advantage of the fact that operators acting on the auxilliary and physical degrees of freedom commute with one another. The exponentiated object $H_{\rm phys}-H^\prime_{\rm aux}$ is effectively the Liouvillian superoperator, $\mathcal{L}$, which conventionally is defined by its action on an operator $\hat O$: $\mathcal{L}\hat O=[H,\hat O]$~\cite{lindner2010}. Indeed,
\begin{equation}
    (H_{\rm phys}-H^\prime_{\rm aux})\hat O_{\rm phys}|\psi_0\rangle=([H,\hat O])_{\rm phys}|\psi_0\rangle.
\end{equation}

In order to compute a spatially-dependent correlation function $\Lambda(x,\tau)$ without finite-size effects, we allow a ``window" within an infinite {%\color{blue}
matrix product state} (MPS) to evolve non-uniformly. As the window is time-evolved, the region of non-uniformity will be housed within a light-cone that expands linearly in time, at least for short times. In order to accommodate this, we allow the system to add un-evolved sites to the boundaries, creating a dynamically-expanding window~\cite{phien2013,milsted2013,Zauner2015}. So long as the threshold for adding sites is set sufficiently low, there will be effectively no finite-size dependence. While the present analysis will focus on uniform properties (corresponding to the Fourier $k=0$ component), we show features of our site-resolved technique in Appendix~\ref{sec:ibcs}.

In order to time-evolve the MPS, we sequentially multiply $|\psi_0\rangle$ by the $W^{\rm II}$ approximation to the time-evolution operator~\cite{zalatel2015,PAECKEL2019}, truncating the bond dimension at each step.   We utilize a third-order split-step method \cite{zalatel2015,PAECKEL2019}, and choose a time step $\Delta\tau=0.01/t$. With these short time steps our method is extremely accurate.  Unlike some alternative techniques, there are no challenges here with dynamically expanding the bond dimension \cite{yang2020}.
%This technique faces no difficulties associated with 
%avoids challenges associated with using   
%and allows for the bond dimension to be expanded dynamically, despite the long-range nature of the effective Hamiltonian acting on a purified density matrix~\cite{yang2020}. 
%{\color{blue} Why is it long-ranged?  Shouldn't it just be next-nearest neighbor terms?}
The results shown here use maximum bond dimensions of $\chi=750-1000$. We emphasize that, for the times shown, our MPS simulations are numerically exact.

\section{Results\label{sec:results}}

\subsection{High-frequency properties}
\begin{figure}
    \centering
    \includegraphics[width=\columnwidth]{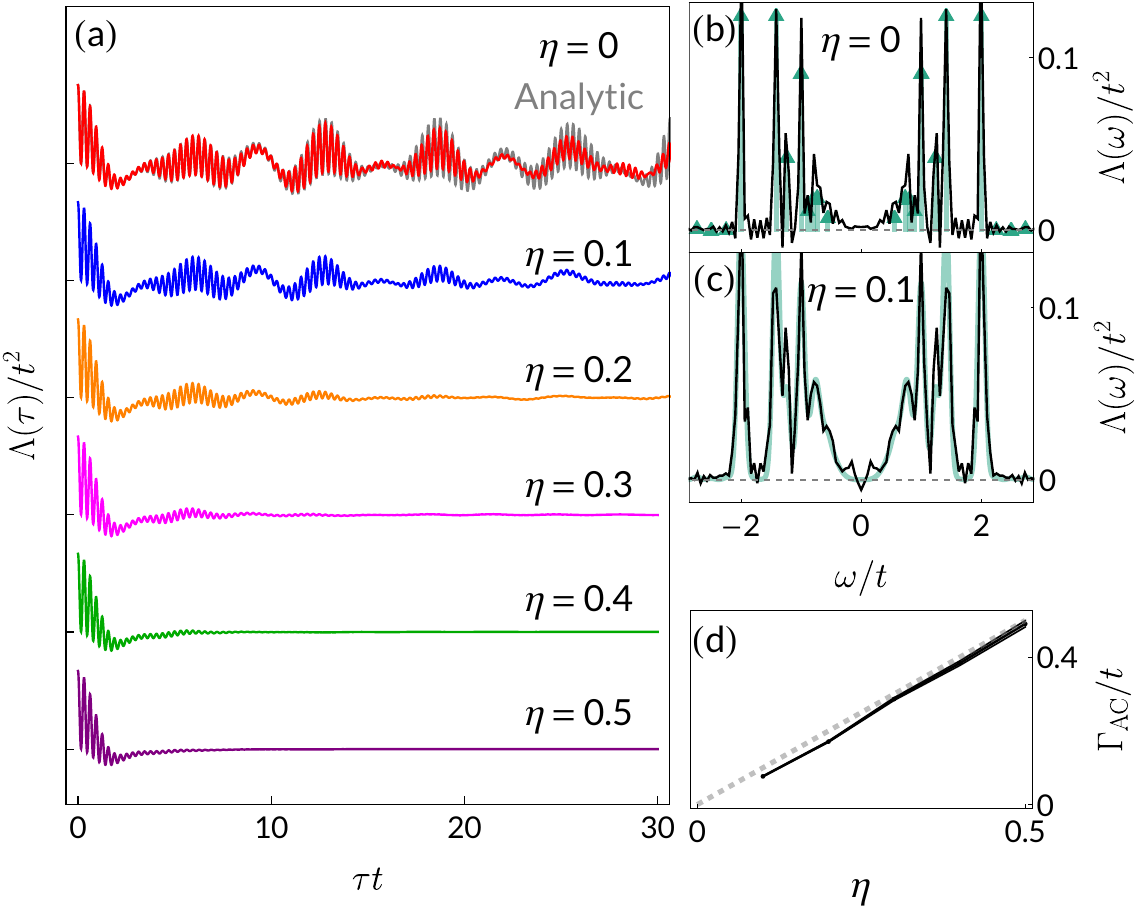}
    \caption{(a) Uniform current-current correlation function, $\Lambda(\tau)=\sum_x\Lambda(x,\tau)$, for $U/t=20$ and a variety of mass ratios. {%\color{blue}
    Curves are offset for visual clarity. Tick marks denote the zero value at each $\eta$, and in all cases $\Lambda(0)=t^2/2$.} For $\eta=0$, there are persistent oscillations out to long times. Our analytic model, derived in the limit $\eta,t/U\to 0$, is shown in gray behind the $\eta=0$ data -- we capture the period of oscillations and their amplitude modulations very accurately. Deviations are due to non-zero $t/U$.  For $\eta>0$, these oscillations are damped by the motion of heavy particles. (b-c) We can extract the damping rate for a given $\eta$ by fitting the Fourier transform of $\Lambda(\tau)$ to the analytic result with a Gaussian broadening factor (see main text). Panel (b) shows the precise agreement for $\eta=0$, without broadening, and panel (c) shows the agreement at $\eta=0.1$. (d) Fitted Gaussian broadening factor, $\Gamma_{\rm AC}$, as a function of $\eta$. Dashed line is $\Gamma_{\rm AC}=\eta$, which appears to describe the data very well.}
    \label{fig:timeseriesFig}
\end{figure}

In Fig.~\ref{fig:timeseriesFig}(a) we show the uniform current-current correlator in the temporal domain, $\Lambda(\tau)=\sum_x\Lambda(x,\tau)$, for $U/t=20$ and a series of mass ratios $\eta=0,~0.1,~0.2,~0.3,~0.4,~0.5$. For small $\eta$, the correlation function is dominated at short times by large oscillations at a frequency $\omega\sim U$, and slower oscillations with $\omega\sim t$. 
The rapid oscillations have an envelope which is modulated at a lower frequency $\omega\sim t$.  %Additionally, there are slow There are also o, corresponding to the envelope width, and its center.  
%at a frequency $\omega\sim t$.  These oscillations are superimposed on oscillations in the mean value of the correlator that also occur at frequencies $\omega\sim t$.
{%\color{blue} 
Note that $\Lambda(0)=t^2/2$ is constrained by a sum-rule and is independent of $\eta$.}

For infinite mass imbalances ($\eta=0$), each of these oscillatory components persist indefinitely. This can be understood by recognizing that the Falicov-Kimball model is an exactly-solvable model of free fermions in a binary-disordered background potential. In one dimension, this system is Anderson-localized for arbitrarily small $U/t$, so $\sigma(\omega=0)=0$. For large $U/t$, the light-particle wavefunctions are all localized to regions of constant background potential (i.e. regions in which $\langle n_{\downarrow}\rangle=0$ or 1 throughout). %Hence, w
When we take the thermal ensemble average, the properties of the system can be written as a sum of contributions from disjoint regions, each of which possess a quantized energy spectrum.
%, each from a distinct region.  
%in terms of the properties of a particle-in-a-box. 
%This implies that t
Consequently, the Fourier transform of the current-current correlator, $\Lambda(\omega)$, is %simply 
a discrete sum over 
%a series of finite-frequency 
delta functions. In Appendix~\ref{sec:analyticDeets}, we carry out an analytic calculation of $\Lambda(\tau)$ when the single-particle wavefunctions are completely localized, i.e. the limit $\eta,t/U\to 0$. The resulting time series is plotted as the gray curve in Fig.~\ref{fig:timeseriesFig}a, sitting behind the $\eta=0$ data, which closely matches the numerical results out past 10 tunneling times. Deviations from the analytic expression are due to the finite localization length at $U/t=20$.

As shown in Fig.~\ref{fig:timeseriesFig}, the long-lived oscillations at $\eta=0$ are damped for finite $\eta$. 
In the frequency domain, we find that the finite $\eta$ spectra are well approximated by Gaussian broadening our analytic $\eta=0$ results.  Figure~\ref{fig:timeseriesFig}b shows the Fourier transform of the $\eta=0$ data from panel $(a)$.  The resulting peaks are well-aligned with the locations of the delta functions in our analytic theory, shown with colored triangles.  Figure ~\ref{fig:timeseriesFig}{%\color{blue}
c} shows the spectrum for $\eta=0.1$.  Colored lines depict our analytic $\eta=0$ result, broadened by a Gaussian of width $\Gamma_{\rm AC}/t=0.077$.  For each value of $\eta$, we find the best fit $\Gamma_{\rm AC}$. 
%We extract the damping rate for these finite-frequency resonances, $\Gamma_{\rm AC}$, by fitting the numerically-obtained $\Lambda(\omega)$ to the analytic result with a Gaussian broadening factor. Fig.~\ref{fig:timeseriesFig}b shows how the $\eta=0$ result aligns closely with the expected delta function peaks, and Fig.~\ref{fig:timeseriesFig}b shows a broadened fit for the $\eta=0.1$ data. We plot the extracted $\Gamma_{\rm AC}$ versus $\eta$ in 
As shown in
Fig.~\ref{fig:timeseriesFig}d,
$\Gamma_{\rm AC}$ is %manifestly 
proportional to $\eta$ with a proportionality constant close to 1 (gray dashed line). This provides a strong indication that the principle damping mechanism is the motion of heavy particles, which should occur on timescales $\sim1/\eta t$, hence resulting in $\Gamma_{\rm AC}\propto\eta$. While this AC damping rate is not a common transport coefficient to measure in condensed matter systems, cold atom experiments are well-placed to extract it by studying the envelope of $\omega\sim U$ oscillations in the current-current correlation functions (see Sec.~\ref{sec:experiment}).

% Figure ()b plots $\Lambda(\tau)=\sum_x\Lambda(x,\tau)$, the $k=0$ Fourier component of the real-time current-current correlator. Here we see long-lived oscillations for $\eta=0$, while $\Lambda(\tau)$ is damped for finite $\eta$. Moreover, the features on the scale $1/U$ are well-separated from those that oscillate as $1/t$. We can make this separation sharper by Fourier transforming again, obtaining $\tilde\Lambda(\omega)$, which is plotted in Fig. ()c for $\eta=0$. Spectral weight is localized to the vicinity of $\omega=0$ and $\omega=U$ within a range of $\Delta\omega\sim4t$. Shaded arrows indicate the location of delta function peaks, extracted from the analytic expansion about the $\eta,1/U\to 0$ limit (see Sec.~\ref{sec:analytic} for more details). We can extract an approximate transport decay coefficient by fitting $\tilde\Lambda(\omega)$ at finite $\eta$ to the analytic resonances, broadened by a decay coefficient $\Gamma$.

\subsection{Low-frequency properties}
\begin{figure}
    \centering
    \includegraphics[width=\columnwidth]{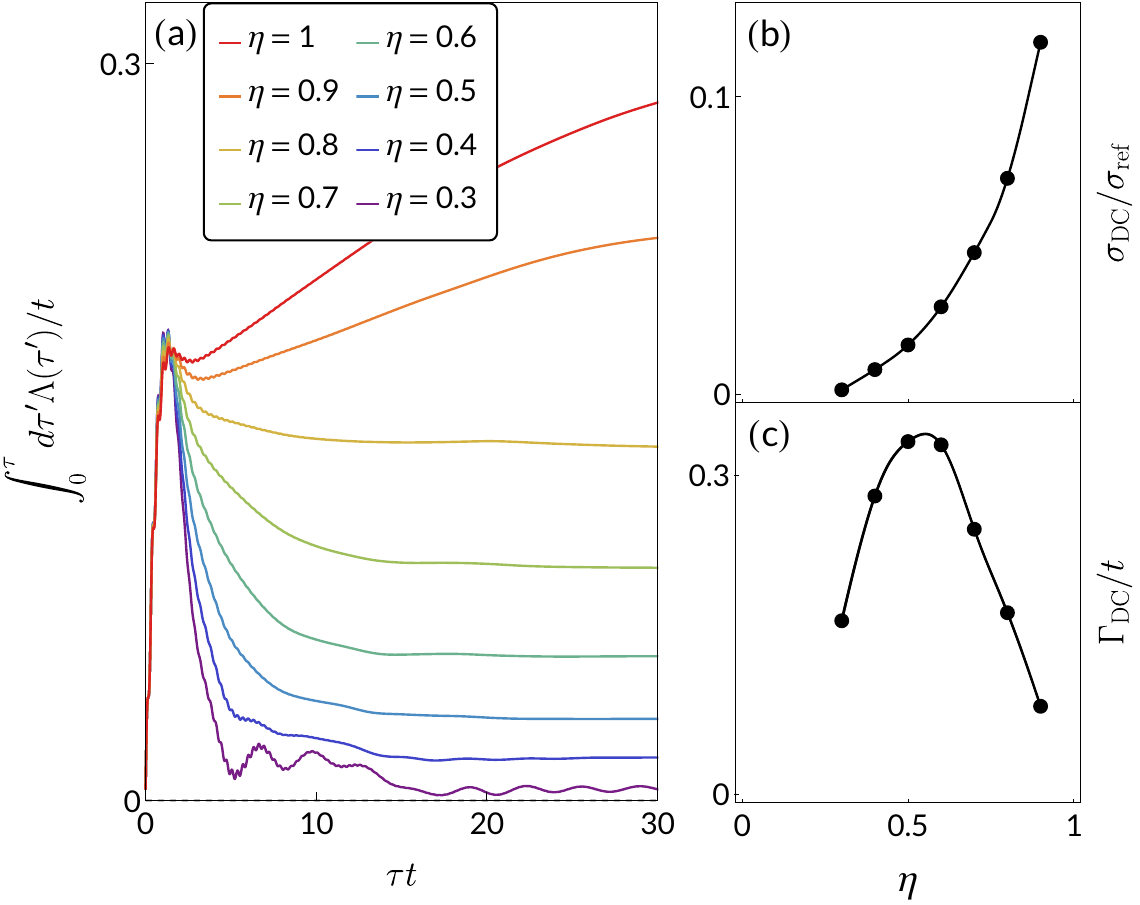}
    \caption{(a) Integral of the uniform current-current correlation function, $\int_0^\tau d\tau^\prime\Lambda(\tau^\prime)$, for $U/t=20$ and a variety of mass ratios. Integrating the timeseries reduces the contributions from rapidly oscillating terms, allowing one to more easily extract the low-frequency behavior. For $\eta=0.3$, oscillations with $\omega\sim t$ have a substantial amplitude even out to 30 tunneling times; these oscillations are even larger for $\eta<0.3$, so those curves have been omitted for visual clarity. Fitting the long-time behavior of these curves to the form $T\sigma_{\rm DC}/2+Ce^{-\Gamma_{\rm DC}\tau}$, we can extract the DC conductivity, $\sigma_{\rm DC}$ and the DC transport relaxation rate, $\Gamma_{\rm DC}$. These are shown as a function of $\eta$ in panels (b) and (c), respectively. We find that $\sigma_{\rm DC}$ appears to vanish continuously as $\eta\to 0$ and diverges as $\eta\to 1$. The transport relaxation rate exhibits non-monotonic behavior, peaking around $\eta=0.5$ and vanishing at $\eta=0,1$. These features are consistent with the fact that both the $\eta=0$ and $\eta=1$ limits are integrable.}
    \label{fig:lg_timeseriesFig}
\end{figure}

In Fig.~\ref{fig:lg_timeseriesFig}(a) we show the integrated uniform current-current correlator, $\int_0^\tau d\tau^\prime\Lambda(\tau^\prime)$, for a variety of mass ratios. This quantity is convenient for studying the low-frequency properties of $\Lambda(\tau)$, as contributions from components oscillating with frequency $\omega$ will generically be diminished by a factor of $1/\omega^2$. For small $\eta$, however, these components are nonetheless substantial, and hence those curves are omitted from Fig.~\ref{fig:lg_timeseriesFig} for clarity. 

After a short-time increase, the integrated timeseries for intermediate values of $\eta$ slowly relax to a constant value at long times. Up to prefactors, this asymptotic value is the DC conductivity: $\lim_{\tau\to\infty}\int_0^\tau d\tau^\prime\Lambda(\tau^\prime)/t=2\sigma_{\rm DC}/\sigma_{\rm ref}$ (c.f. Eq.~(\ref{eq:highTSigma})). 
As introduced in Sec.~\ref{sec:formalism}, $\sigma_{\rm ref}=a\beta t$ where $a$ is the lattice spacing, $\beta$ the inverse temperature, and $t$ the hopping matrix element.
We extract the asymptotic conductivity, $\sigma_{\rm DC}$ as well as the transport relaxation rate, $\Gamma_{\rm DC}$, by fitting the integrated timeseries to the form $\int_0^\tau d\tau^\prime\Lambda(\tau^\prime)/t=2\sigma_{\rm DC}/\sigma_{\rm ref}+Ce^{-\Gamma_{\rm DC}\tau}$ with $\{C,\sigma_{\rm DC},\Gamma_{\rm DC}\}$ as fitting parameters. {%\color{blue} 
Fitting was performed for $\tau\geq 6/t$.} These two DC transport properties are shown as functions of $\eta$ in Fig.~\ref{fig:lg_timeseriesFig}b and c, respectively.

We find that the DC conductivity diverges continuously as $\eta\to 1$, while the relaxation rate vanishes. This divergence is consistent with %the %integrable symmetric limit as described earlier: the model is expected to display 
%divergent conductivity expected in the integrable limit,
%$\eta=1$.  Here one expects 
%unconventional KPZ dynamical scaling, which manifests as 
superdiffusive behavior at $\eta=1$.
%in the current-current correlator %Superdiffusivity is associated with an integrated correlator that asymptotically scales as $\int_0^\tau d\tau^\prime\Lambda(\tau^\prime)/t\propto \tau^\alpha$ with 
%${0<\alpha<1}$.  This behavior is related to a low-frequency power-law divergence $\tilde\Lambda(\omega)\propto|\omega|^{-(1+\alpha)}$. Indeed, we find that the integrated timeseries at $\eta=1$ does not appear to plateau at any finite time, nor does it diverge linearly with time. Our finite-time results provide only a crude estimate of the superdiffusive exponent, 
% {\color{blue} 
Superdiffusivity is characterized by a dynamical exponent $1<z<2$, which describes the long-time hydrodynamic relationship between spatial and temporal fluctuations; diffusive transport corresponds to $z=2$, while ballistic transport corresponds to $z=1$. For the half-filled 1D Hubbard model one expects KPZ dynamical scaling, corresponding to a dynamical exponent $z=3/2$~\cite{fava2020}. As described in Appendix~\ref{sec:kpz}, the dynamical exponent is directly related to the long-time behavior of the integrated current-current correlator: $\int_0^\tau d\tau^\prime\Lambda(\tau^\prime)/t\propto \tau^\alpha$ with $\alpha=2/z-1$~\cite{bertini2021}. 
This implies a power-law divergence in the optical conductivity, $\sigma(\omega)\propto |\omega|^{-\alpha}$. Reliably extracting $\alpha$ from our data is challenging, as we have a relatively short time window over which we can fit the power law:  Short-time transients persist to $\tau\sim 10/t$, and our
time series only extends to $\tau=30/t$.  Our best fit yields
%only have half a decade of data, our finite-time results %provide only a crude estimate of the superdiffusive exponent, 
%suggest
$\alpha\sim0.3$, with errors that are dominated by systematics and are therefore hard to quantify. This result compares reasonably well with the expected value of $\alpha=1/3$.
%using the expected KPZ dynamical exponent $z=3/2$. %}
%$0.2\lesssim\alpha\lesssim0.3$; we have found no prior theoretical predictions to which this exponent can be compared. 

In the opposite limit, $\eta\to0$, we find that the conductivity and the relaxation time appear to continuously vanish. This is consistent with an Anderson-localized infinite-temperature state at $\eta=0$, in which the effective free-particle excitations do not relax. For $\eta<0.3$, we are unable to reliably extract 
$\sigma_{\rm DC}$ or $\Gamma_{\rm DC}$
%these coefficients 
from the finite-time current correlations due to large and persistent oscillations in the integrated correlator.

\subsection{Interaction dependence}

\begin{figure}
    \centering
    \includegraphics[width=\columnwidth]{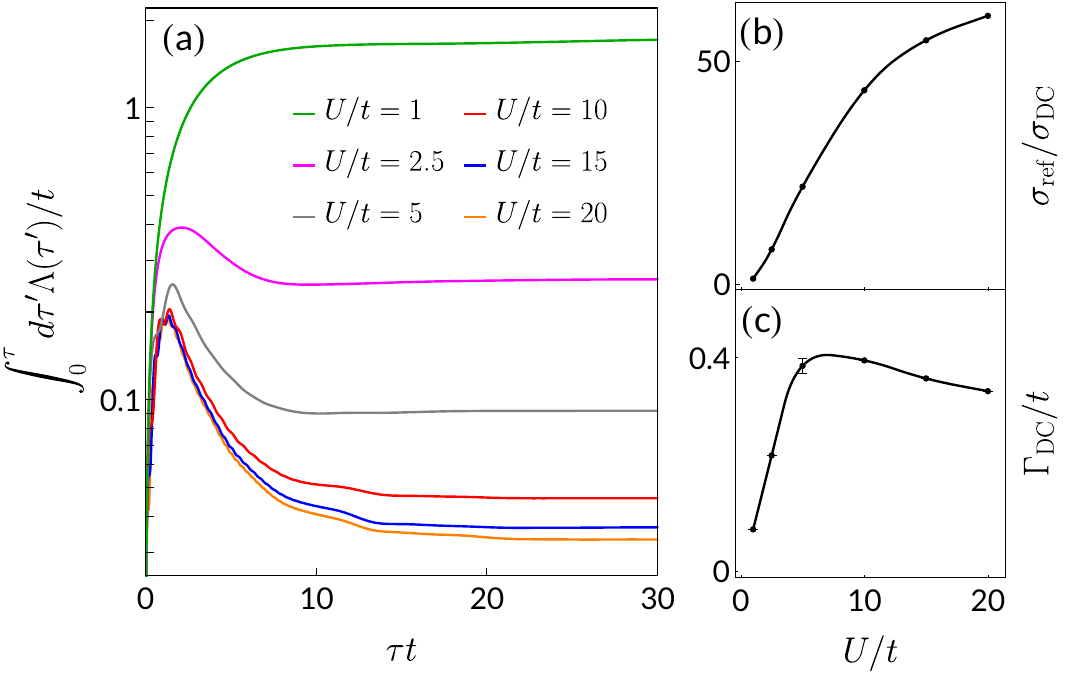}
    \caption{(a) Integral of $\Lambda(\tau)$ out to finite times for $\eta=0.5$ and various values of $U/t$. Note the log scale on the y-axis. (b) DC resistivity (inverse of the conductivity) versus $U/t$ {%\color{blue} 
    from a fit to (a)}. The resistivity vanishes continuously as we approach the non-interacting limit, $U/t\to 0$. As $U/t\to\infty$, the resistivity is expected to plateau at a finite value. (c) Transport decay rate versus $U/t$ {%\color{blue} 
    from a fit to (a)}. The decay rate vanishes as $U/t\to 0$ and appears to peak around $U\sim 8t$. As $U/t\to\infty$, $\Gamma_{\rm DC}$ approaches a finite, $\eta$-dependent value.}
    \label{fig:udependence}
\end{figure}

In Fig.~\ref{fig:udependence}a, we show the integrated timeseries for a variety of $U/t=1,2.5,5,10,15,20$ at fixed $\eta=0.5$. In this non-integrable limit, the relaxation timescales are relatively short and our numerics %should 
provide a reasonable estimate for the interaction-dependence of the DC transport properties. Fig.~\ref{fig:udependence}b shows the resistivity, $1/\sigma_{\rm DC}$, as a function of $U/t$. The resistivity vanishes continuously in the weakly-interacting limit, $U/t\to 0$. Second-order perturbation theory~\cite{kiely2021,zechmann2022} suggests that the resistivity should vanish $\propto (U/t)^2$ in this limit, %the limit of weak interactions, 
which is consistent with our results. For large $U/t$, we expect the resistivity to plateau at a finite value, though at $U/t=20$ it has not yet saturated. Note that this large-interaction limit does not approach a Mott insulator (which would have an infinite resistivity at half-filling) because we took the limit of infinite temperature first. 

As shown in Fig.~\ref{fig:udependence}c, the transport relaxation rate $\Gamma_{\rm DC}$ displays similar behavior to the resistivity: it vanishes as $U/t\to 0$ and plateaus as $U/t\to\infty$. At weak coupling, one should find a Drude-like relationship between the resistivity and the scattering rate: $1/\sigma_{\rm DC}\propto\left(m^*/n\right)_{\rm eff}\Gamma_{\rm DC}$. This suggests that $\Gamma_{\rm DC}\propto U^2/t$, although we are not able to resolve this feature. 
%in Fig.~\ref{fig:udependence}c. 
In the strong-coupling limit, one generically expects the transport scattering rate to saturate at a constant value set by the lattice spacing~\cite{kiely2021}.

\section{Analytic Expansion\label{sec:analytic}}
In the limit $\eta\to0$ the heavy particles become frozen, and the interaction term in the Hamiltonian becomes a static disorder potential.  The resulting non-interacting model is much simpler than the original.  Here we develop an analytic expansion which lets us calculate $\Lambda(\tau)$ for $\eta=0$ when $U\gg t$.

%us all properties can be calculated in terms of a non-interacting

%Calculating the properties of the light particles then reduces to solving 
%In this limit where the heavy particles cannot hop, their configuration is given by specifying the occupations $\{n_{i,\downarrow}\}$, where $n_{i,\downarrow}=0,1$ for each site $i$.  

For a given configuration  of the heavy particles, $\{n\}=\{ n_{i,\downarrow}\}$, the motion of the light particles is controlled by a Hamiltonian
\begin{multline}
    \mathcal{H}_{0}=-t\sum_{i} \left(c^\dagger_{i,\uparrow}c_{i+1,\uparrow}+{\rm H.c.}\right)+\sum_iV(r_i)n_{i,\uparrow}
    \label{eq:h0_quadratic_main}
\end{multline}
where $V(r_i)=U\langle n_{i,\downarrow}\rangle$.
We can use Wick's theorem to write the current-current correlator of this model in terms of the single particle Greens functions, $G^>_{ij}(\tau)=\langle c_i(\tau) c_j^\dagger(0)\rangle$ and $G^<_{ij}(\tau)=-\langle c_j^\dagger(0) c_i(\tau)\rangle$:
% \begin{equation}\label{slam}
% \Lambda_{\{n\}}(x,\tau)=-\frac{t^2}{N_s} \sum_{i,s_1 s_2}
% s_1 s_2\, G^>_{i,i+x}(\tau)
% G^<_{i+x+s_1,i+s_2}(-\tau).
% \end{equation}
\begin{equation}
\Lambda_{\{n\}}(x,\tau)=-\frac{t^2}{N_s} \sum_{\substack{{i,j,k,l}\\j-i=x}}
\eta_{ik}\eta_{jl}\, G^>_{i,j}(\tau)
G^<_{k,l}(-\tau)\label{slam}
\end{equation}
where
\begin{equation}
    \eta_{ik}=
    \begin{cases}
        1 & {\rm if}~k-i=1\\
        -1 & {\rm if}~k-i=-1\\
        0 & {\rm otherwise}
    \end{cases}
\end{equation}
 and $i,j,k,l$ are summed over all $N_s$ lattice sites with the constraint that $j-i=x$.
% Here $s=\pm 1$ and we used time-translation invariance. 
We then calculate $\Lambda(x,\tau)$ by performing a disorder average over $\Lambda_{\{n\}}(x,\tau)$.  

The potential in Eq.~(\ref{eq:h0_quadratic_main}) breaks the lattice into a series of disjoint regions over which $V(r_i)$ is constant (either 0 or $U$).  We are interested in the large $U$ limit:  To leading order,
%For infinite $U$, 
a light particle which is placed in one of these regions cannot leave -- the energy eigenstates are localized to these regions.  Thus
the Greens functions $G_{ij}$ vanish unless $i$ and $j$ are in the same region.  Equation~(\ref{slam}) can then be written as the sum of two terms: $\Lambda^{(0)}$, for which both Greens function describe motion in the same region, and $\Lambda^{(U)}$ where they are in neighboring regions.  The former gives dynamics on the timescale $1/t$, while the latter gives dynamics on the scale $1/U$.

As detailed in Appendix~\ref{sec:analyticDeets}, the resulting spectrum of $\Lambda$ consists of a sum of delta-function peaks, illustrated in Fig.~\ref{fig:timeseriesFig}b.  The low-frequency peaks are at energies $\omega= 2t(\cos \pi n/(m+1) - \cos \pi n^\prime/(m+1))$ where $n,n^\prime,m$ are integers with distinct parity (ie. $n\pm n^\prime$ is odd) with $1\leq n,n^\prime\leq m$.  The parity constraint comes from the symmetry of the current operator, and implies that there is no peak at $\omega=0$.  Here $m$ corresponds to the length of the contributing region.  Large-$m$ regions are exponentially suppressed, and the dominant peaks come from $m=2$ and $m=3$.  This results in the ``U" shaped distribution of spectral weight, as seen in Fig~\ref{fig:timeseriesFig}b.  The high-frequency peaks are at energies $\omega=\pm U+ 2t(\cos \pi n/(m+1) - \cos\pi n^\prime/(m^\prime+1))$ where $n,n^\prime,m,$ and $m^\prime$ are integers with $1\leq n\leq m$ and $1\leq n^\prime\leq m^\prime$.  There are no parity constraints and $m,m^\prime$ represent the lengths of neighboring regions.  The dominant term comes from $n=n^\prime$, resulting in $\omega=\pm U$.  Again, the large $m,m^\prime$ terms are exponentially small.
%-- resulting in the observed peak-shape.

The analytic peak locations and amplitudes appear to match well the numerical results from our MPS calculations.  The broadening of the black curve in Fig.~\ref{fig:timeseriesFig}b comes from the finite time-window of our numerical data.

\section{Experimental Implementation\label{sec:experiment}}
Transport measurements in ultracold atomic systems are in general quite challenging to implement. Recent experiments have managed to measure the DC conductivity via the Einstein relation~\cite{Brown2019}, the optical conductivity via a modulated trap potential~\cite{thywissen2019}, and the momentum relaxation rate~\cite{demarco2019}. Here, we propose an alternative technique to measure the optical conductivity in the \textit{time} domain, and hence $\Lambda(\tau)$ at high temperatures. In this way, our numerical results can be studied directly using the present generation of ultracold atom experiments.

\begin{figure}
    \centering
    \includegraphics[width=\columnwidth]{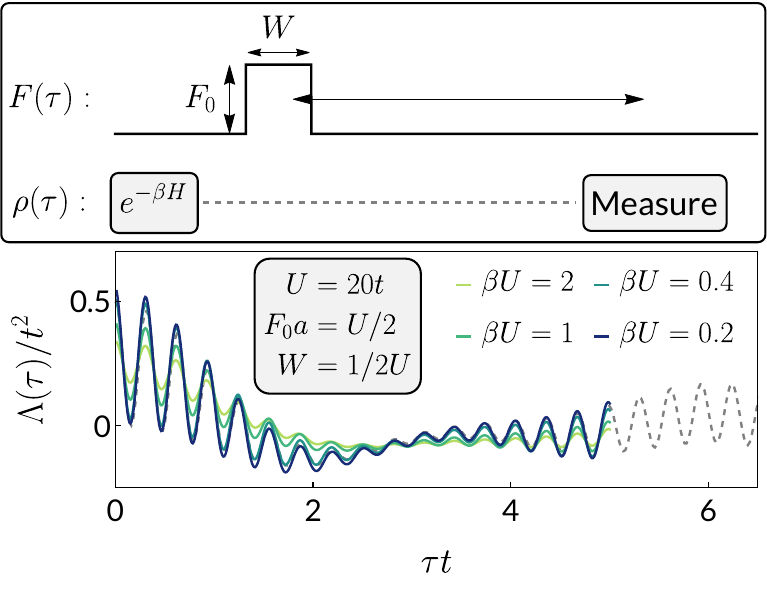}
    \caption{Experimental sequence for measuring current-current correlations in an ultracold atomic gas. (top) After preparing a thermal ensemble, one pulses on a lattice tilt for time $T$. One then allows the system to evolve for time $\tau$ and measures the momentum distribution using time-of-flight (see main text). (bottom) Benchmarking this procedure using the TDVP algorithm on a 30-site lattice using Eq.~(\ref{eq:lambdaJ}). {%\color{blue} 
    Solid curves show TDVP calculations for various parameters.} Simulations approach the infinite-system, {%\color{blue}
    infinite-temperature} results {%\color{blue}
    (gray dashed line)} when the parameters $F_0a/U$, $WU$ and $\beta U$ are made small. {%\color{blue} 
    Dashed curve is continued to longer times for visual clarity.}}
    \label{fig:experiment}
\end{figure}
Our experimental procedure is schematically illustrated in Fig.~\ref{fig:experiment}. After preparing a thermal ensemble, we propose pulsing on a lattice tilt, $H_{\rm tilt}(\tau)=F(\tau)\sum_l ln_l$, for a short time, $W$. In quantum gas microscopes, this linear potential can be realized 
from the AC stark shift at the edge of
a wide Gaussian beam~\cite{guardadoSanchez2020}. Generically, this pulsed tilt will generate a current response given by $\langle J\rangle(\beta,\omega) = \sigma(\beta,\omega)\tilde F(\omega)$ where $\tilde F(\omega)$ is the Fourier transform of $F(\tau)$ and $\sigma(\beta,\omega)$ is the optical conductivity at inverse temperature $\beta$. In the limit that the pulse duration $W\to 0$, we can approximate $F(\tau)\approx F_0W\delta(\tau)$, and hence the real-time current response is given by $\langle J\rangle(\beta,\tau)=F_0W\sigma(\beta,\tau)$. In the high-temperature $\beta\to 0$ limit, then, we use Eq.~(\ref{eq:highTSigma}) to find
\begin{equation}
\langle J\rangle(\beta,\tau)\approx\frac{\beta F_0W}{2} \Lambda(\tau).
\label{eq:lambdaJ}
\end{equation}
In quantum gas microscopes, the expectation value of the current operator can be measured by releasing the atoms from their trapping potential, allowing them to expand and thus mapping their \textit{in situ} momenta to position space~\cite{bloch2008,chin2010}. Once the momentum distribution function {%\color{blue}
$\langle n_{k,\mu}\rangle$ is known, the expectation value of the current is simply $\langle J\rangle=\sum_{k,\mu}t_\mu\sin(k)\langle n_{k,\mu}\rangle$.}

In the bottom panel of Fig.~\ref{fig:experiment}, we show the results of this procedure simulated with the finite-size TDVP (time-dependent variational principle) algorithm~\cite{haegeman2011} on a $30$-site lattice for $U/t=20$ and various values of $\beta$. 
This modeling captures experimental finite-size effects, as well as the influence of detailed pulse shapes.
%Importantly, as our procedure picks out a signal proportional to $\beta$, we cannot take the true infinite-temperature limit here. 
As noted above, our results emerge in the $\beta\to 0$ limit; finite-$\beta$ corrections  yield additional information about the full optical conductivity $\sigma(\beta,\tau)$. We find that the results agree well with our infinite calculations so long as $\beta U$, $W U$, and $F_0a/U$ are all $\lesssim 1$ (assuming $U\gg t$).

{%\color{blue}
This proposed experimental procedure yields the conductivity at any temperature.  Comparison with our theoretical results, however, is predicated on 
the simultaneous validity of the single band approximation and the high temperature limit.  This  requires $W\ll T\ll\Delta$, where $\Delta$ is the band gap and $W$ is the band width.   Both $W$ and $\Delta$ can be estimated by solving the single particle Schrodinger equation in a sinusoidal potential $V(x)=V_0 \cos^2(k_R x)$.  The Ytterbium experiment~\cite{darkwah2022} was performed with $V_0=7 E_R$, where $E_R=\hbar^2 k_R^2/2m$. This leads to $W/\Delta\approx 0.05$, indicating that there is a temperature regime where our single band high temperature expansion is relevant.  
%In our previous work we found that at weak coupling the leading term in the high temperature expansion dominates transport down to surprisingly low temperatures \cite{kiely2021}. 
%This ratio  falls exponentially with lattice depth $V_0$, and hence the separation of scales can readily be made even larger. 
%Thus our high temperature limit is experimentally achievable.
%In ultracold atom systems, the single-band Fermi-Hubbard Hamiltonian emerges as an effective low-energy description, on scales small compared to band gap.  For a lattice dept of $V_0=10 E_R$, the band gap is roughly 60 times larger than the band width
%the recoil energy, $E_R/h\sim2$~kHz for Ytterbium~\cite{darkwah2022}. For a lattice of depth $V_0=7 E_R$ the hopping is $t/h\sim100$~Hz. 
%This separation of scales 
%These energy scales are separated by 
%Thus our calculations are valid in the regime $2\times10^{-8}~{\rm K}<k_B T<10^{-7}~{\rm K}$.}
% {\color{blue}One might then ask how reasonable the high-temperature expansion really is. Indeed, such an expansion relies on the effective temperature being comparable to the bandwidth and interaction strength but small compared to the band-gap. Previous work~\cite{Brown2019,kiely2021} has found that this regime is straightforwardly achievable in ultracold atom quantum simulators.
}

\section{Conclusions and Outlook\label{sec:discussion}}
In this paper, we have have a provided a comprehensive study of the moderate-time transport properties of the one-dimensional mass-imbalanced Fermi-Hubbard model in the high-temperature limit. This model serves as an valuable setting for exploring the interplay of integrability and localization. 

By studying the strongly-interacting and heavily-imbalanced limit, we have mapped out the resonances associated with bound excitations of heavy and light particles. Our analytic model provides a precise account of these non-trivial finite-frequency features, which we have benchmarked against state of the art time-dependent MPS simulations.
% {\color{blue}
In the strongly-interacting and weakly-imbalanced limit, we found that the DC conductivity diverges continuously as $\eta\to 1$. At the symmetric point, superdiffusive correlations lead to a power-law divergence in the integrated correlator as a function of time.%, $\tau^\alpha$. 
This corresponds to a power-law divergence in the optical conductivity, $\sigma(\omega)\propto|\omega|^{-\alpha}$. We estimate the exponent of this divergence to be $\alpha\sim 0.3$, close to the KPZ prediction of $\alpha=1/3$.
% }

Many %of the 
interesting questions regarding this model remain outstanding. While there is good reason to believe that the mass-imbalanced model is ergodic for $\eta>0$~\cite{deroeck2014b,PAPIC2015714,yao2016,sirker2019}, directly studying the onset of diffusive transport is challenging.
%has not been studied in the strongly-interacting limit. Importantly, q
Questions persist about the long-lived nature of bound states and whether intermediate-scale subdiffusive regimes exist prior to thermalization~\cite{yao2016,darkwah2022}.
% {\color{blue}
Similar questions exist near the symmetric limit: How does the characteristic time for the onset of diffusion, $\tau_{\rm D}$, behave as $\eta\to 1$ in the strongly-correlated limit? 
%What is the expected divergence in the optical conductivity at the symmetric point?
% }
The timescales on which these questions can be answered are not accessible with the current generation of numerical tools, and it will require further theoretical and experimental insights to probe these regimes~\cite{white2018,rakovszky2022,thomas2023comparing}. Our experimental proposal provides one possible approach.
%controlled and benchmarked calculation, as well as our experimental proposal, provides a springboard for future work in pursuit of these goals.

\begin{acknowledgments}
This work was supported by the NSF Grant No. PHY-2110250.
\end{acknowledgments}

\appendix

%\section{Details of iMPS Calculation\label{sec:timeEvol}}

\section{Details of MPS calculation}

\subsection{MPS Purification\label{sec:timeEvol}}

In order to model the properties of the infinite-temperature current-current correlation function, we first purify the %infinite-temperature 
density matrix so that it can be represented as an infinite matrix product state (iMPS). This is a standard technique, so we refer the interested reader to Ref.~\cite{SCHOLLWOCKreview} for a thorough review. In this section, we will give a brief introduction to purification and then describe the procedure that we use for incorporating number conservation.

The infinite-temperature density matrix, $\rho_0$, is simply a diagonal operator acting on the physical Hilbert space of the Hamiltonian:
\begin{equation}
    \rho_0=\prod_i\bigg[\frac{1}{\mathcal{N}_i}\sum_{s_i,s_i^\prime}\delta_{s_i,s^\prime_i}y(s_i)|s_i\rangle\langle s^\prime_i|\bigg]
\end{equation}
where $i$ indexes lattice sites and $s_i,s_i^\prime \in\{\uparrow,\downarrow,\varnothing,\uparrow\downarrow\}$ span the local Hilbert space on site $i$. Importantly, this form of $\rho_0$ shows that distinct sites $i$ and $j$ are completely uncorrelated. The quantity $y(s_i)$ is the fugacity and $\mathcal{N}_i$ is the normalization enforcing ${\rm Tr}\rho_0=1$. At infinite temperature, zero magnetization and a fixed particle density $\bar n$, we have $y(\uparrow)=y(\downarrow)=\bar n/(2-\bar n)$, $y(\varnothing)=1$ and $y(\uparrow\downarrow)=y(\uparrow)y(\downarrow)$, which means $\mathcal{N}_i=(1+y(\uparrow\downarrow))^2$.

We will now introduce an auxiliary set of states, labeled by quantum number  $\{t\}$, which expand our Hilbert space and  allow us to represent the density matrix as  $\rho_0={\rm Tr}_{\{t\}}|\psi_0\rangle\langle\psi_0|$.  The simplest such representation is  
\begin{equation}
    |\psi_0^{\rm triv}\rangle=\prod_i\bigg[ \frac{1}{\sqrt{\mathcal{N}_i}}\sum_{s_i,t_i}\delta_{s_i,t_i}\sqrt{y(s_i)}|s_i\rangle\otimes|t_i\rangle \bigg].
\end{equation}
Here $i$ indexes the physical sites, as before, but now it is associated with \textit{two} sites: the same physical one, whose state is denoted by $s_i$, and the associated auxiliary one, whose state is denoted by $t_i$. We refer to this as a trivial purification because it fixes the state of the physical site $s_i$ to be identical to that of the auxiliary site $t_i$.  

As explained in Sec.~\ref{sec:sym}, it is convenient to 
construct alternative purifications via unitary transformations on the auxiliary subspace.  These will make it easier to encode conservation laws.  Explicitly, 
{%\color{blue}
\begin{equation}
\begin{split}
    |\psi_0(U)\rangle&= U^\dagger|\psi_0^{\rm triv}\rangle \\
    &=\prod_i\bigg[ \frac{1}{\sqrt{\mathcal{N}_i}}\sum_{s_i,t_i}\delta_{s_i,t_i}U^*_{\tilde t_i,t_i}\sqrt{y(s_i)}|s_i\rangle\otimes|\tilde t_i\rangle \bigg] \\
    &=\prod_i\bigg[ \frac{1}{\sqrt{\mathcal{N}_i}}\sum_{s_i,t_i}U^*_{t_i,s_i}\sqrt{y(s_i)}|s_i\rangle\otimes|t_i\rangle \bigg].
\end{split}
\end{equation}}
For arbitrary unitary transformations, the infinite-temperature density matrix can be written as a partial trace over the auxiliary degrees of freedom:
{%\color{blue}
\begin{equation}\label{rhorep}
\begin{split}
    \rho_0&={\rm Tr}_{\{t,t^\prime\}}|\psi_0(U)\rangle\langle\psi_0(U)| \\
    &=\prod_i\bigg[\frac{1}{\mathcal{N}_i}\sum_{s_i,s_i^\prime,t_i,t_i^\prime}\sqrt{y(s_i)y(s_i^\prime)}U^*_{t_i,s_i}|s_i\rangle\langle t_i|t_i^\prime\rangle\langle s^\prime_i|U_{t_i^\prime,s_i^\prime}\bigg] \\
    &=\prod_i\bigg[\frac{1}{\mathcal{N}_i}\sum_{s_i,s_i^\prime}\delta_{s_i,s^\prime_i}y(s_i)|s_i\rangle\langle s^\prime_i|\bigg].
\end{split}
\end{equation} }

\subsection{Symmetries\label{sec:sym}}

%The unitary degree of freedom in Eq.~(\ref{rhorep}) can be very useful for algorithmic efficiency,  aiding in producing a representation for which one can take advantage of symmetries.

%note that the trivial purification $|\psi_0^{\rm triv}\rangle$ can be represented as a MPS with a two-site unit cell with alternating bond dimensions equal to $1$ and ${\rm dim}(s_i)$ (which is equal to the dimension of the local Hilbert space). Such a state is very efficient to construct, although it explicitly breaks all symmetries of the underlying model. This is problematic, as symmetry-conserving MPS wavefunctions can be written in a block-sparse form~\cite{SCHOLLWOCKreview,kiely2022} that leads to an algorithmic speedup of all tensor operations.

One often uses conservation laws to write MPS tensors in a block-sparse form, speeding up the calculations~\cite{SCHOLLWOCKreview,kiely2022}.  
In our case, the Fermi-Hubbard model conserves both total magnetization and total particle number: ${\rm U}_{\rm part}(1)\otimes{\rm U}_{\rm spin}(1)$. For the wavefunction on the doubled Hilbert space, then, we should double the symmetry: ${\rm U}^{\rm phys}_{\rm part}(1)\otimes{\rm U}_{\rm part}^{\rm aux}(1)\otimes{\rm U}^{\rm phys}_{\rm spin}(1)\otimes{\rm U}^{\rm aux}_{\rm spin}(1)$. Unfortunately, total magnetization and particle number are explicitly not conserved for $|\psi_0(U)\rangle$, as it represents a  purified grand-canonical density matrix.
Working in the canonical ensemble is intractable, as that leads to a MPS whose bond dimension scales with the system size~\cite{kiely2022}.  Nonetheless, we can take advantage of some of the symmetry: In the trivial purification, for each spin state the number of auxiliary particles equals the number of physical particles.

A convenient way to keep track of this symmetry is to introduce a particle-hole transformation on the auxiliary particles,
%, so the bond dimension required to conserve the full $[{\rm U}(1)]^4$ symmetry diverges with the length of the system~\cite{kiely2022}. Hence, such a wavefunction is intractable in the thermodynamic limit. With that said, a simple choice of $U$ can be used to conserve the \textit{relative} magnetization and particle number, leading to a significant reduction in computational resources. Specifically, we take 
$U=\prod_i U_i$ where
\begin{equation}
    U_i=\begin{pmatrix} 
0 & 1 & 0 & 0 \\
-1 & 0 & 0 & 0 \\
0 & 0 & 0 & -1 \\
0 & 0 & 1 & 0 \\
\end{pmatrix}\label{eq:ourunitary}
\end{equation}
is written in the basis introduced above: $\begin{pmatrix}\uparrow,\downarrow,\varnothing,\uparrow\downarrow\end{pmatrix}$.
%In other words, our 
This results in a
purified state %takes 
of the form
{%\color{blue}
\begin{multline}\label{eq:purification}
    |\psi_0(U)\rangle=\prod_i\bigg[\frac{1}{\sqrt{\mathcal{N}_i}}\big(-\sqrt{y(\uparrow\downarrow)}|\uparrow\downarrow\rangle_{\rm phys}|\varnothing\rangle_{\rm aux} \\ +\sqrt{y(\varnothing)}|\varnothing\rangle_{\rm phys}|\uparrow\downarrow\rangle_{\rm aux}
    - \sqrt{y(\uparrow)}|\uparrow\rangle_{\rm phys}|\downarrow\rangle_{\rm aux} \\+\sqrt{y(\downarrow)}|\downarrow\rangle_{\rm phys}|\uparrow\rangle_{\rm aux} \big)\bigg]_i.
\end{multline} }
As one sees explicitly, in this representation the total  number of particles (physical plus auxilliary), and their net spin, is fixed.  Hence we can introduce the associated quantum numbers, and all tensors in our MPS calculation will have block-sparse forms.  The minus signs are chosen to simplify the construction in Sec.~\ref{sec:aux}.

\subsection{Auxiliary Operators\label{sec:aux}}
The fact that $\rho_0$ is diagonal means that its purification has a special property: any operator $\hat O_{\rm phys}$ that acts on the physical degrees of freedom of $|\psi_0(U)\rangle$ has a partner, $\hat O_{\rm aux}^\prime$, acting on just the auxiliary degrees of freedom, such that~\cite{KENNES2016}
\begin{equation}
    \hat O_{\rm phys}\otimes\mathbb{I}_{\rm aux}|\psi_0(U)\rangle=\mathbb{I}_{\rm phys}\otimes\hat O_{\rm aux}^\prime|\psi_0(U)\rangle.
\end{equation}
The relationship between $\hat O$ and $\hat O^\prime$ depends on the choice of purification, $|\psi_0(U)\rangle$: $\hat O^\prime = U^\dag \hat O U$. As argued in the main text, determination of the auxiliary Hamiltonian can significantly decrease the computational complexity of time evolution by allowing one to shift the purification insertion point~\cite{PAECKEL2019}. In our case, the choice of unitary in Eq.~(\ref{eq:ourunitary}) implies that the auxiliary Hamiltonian ({%\color{blue} 
at half-filling}) is simply the particle-hole transform of the physical one:
\begin{equation}
\begin{split}
    U^\dagger \hat c_\mu U&=-\hat c^\dagger_\mu \\
    U^\dagger \hat n_\mu U&= 1-\hat n_\mu
    \end{split}
\end{equation}

\subsection{Time evolution with infinite boundary conditions\label{sec:ibcs}}

The numerical technique we employ is based on the dynamically-expanding window technique developed in Refs.~\cite{phien2013,milsted2013,Zauner2015}, and we refer interested readers to these references. In this section
%, for completeness, 
we give a high level discussion of a number of details, explaining the major bottlenecks.
%of the algorithm at a high level.

To calculate our response function, we will apply the local current operator $j(0)=-it(c_{1}^\dagger c_0-c_0^\dagger c_1)$ to the infinite matrix product state described by Eq.~(\ref{eq:purification}).  This is a local perturbation that just influences sites $l=0,1$.  As discussed in the main text, we then time-evolve this wavefunction using the Liouvillian, ${\cal L}=H_{\rm phys}-H^\prime_{\rm aux}$, to produce
\begin{equation}
|\psi(0,\tau)\rangle=e^{i\mathcal{L}\tau}j_{0}|\psi_0\rangle.
\end{equation}
At any finite time $\tau$, the matrix product state representation of $|\psi(0,\tau)\rangle$ will differ from $|\psi_0\rangle$, only over a region of size $l_0(\tau)$, centered at the origin.  The perturbed region grows linearly with time, describing a light cone.

In our calculation we start with an infinite matrix product state ansatz where all but $l_0(\tau=0)=6$ of the matrices have the form Eq.~(\ref{eq:purification}).  As described in the main text, we use a third order split step $W^{II}$ time evolution algorithm, where we sweep back and forth through this active region, updating the $l_0$ matrices.  
%As detailed in \cite{phien2013,milsted2013,Zauner2015}, w
We increase $l_0$ when the entanglement entropy between the bulk and boundary sites surpasses $\epsilon_{\rm thresh}=10^{-5}$~\cite{phien2013,milsted2013,Zauner2015}.

\begin{figure}
    \centering
    \includegraphics[width=\columnwidth]{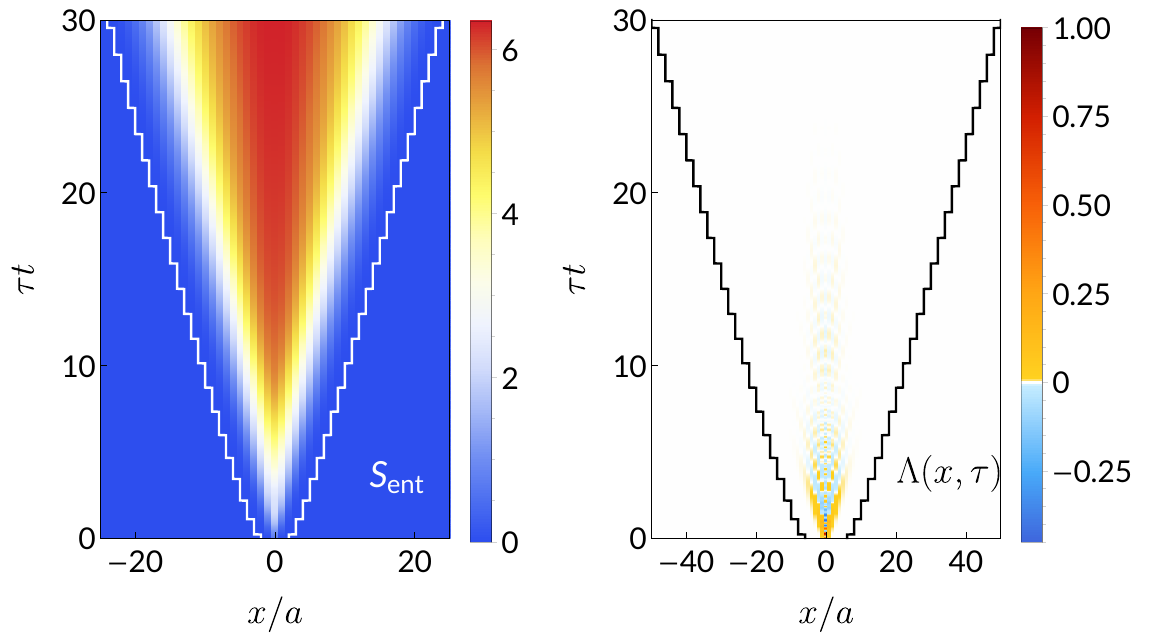}
    \caption{Heat map of the entanglement entropy, $S_{\rm ent}(x,\tau)$ (left), and current-current correlator, $\Lambda(x,\tau)$ (right), as a functions of position and time. On both plots, the effective length of the MPS is shown as a solid jagged line: the white line in (left) is equal to the number of physical sites at a given time, while the black line in (right) is double the number of physical sites (as we take all possible overlaps between two length-$L$ MPS, see Appendix~\ref{sec:ibcs}).}
    \label{fig:ent}
\end{figure}

Figure~\ref{fig:ent}(a) shows the evolution of the entanglement entropy of the purified wavefunction during this process, for $U/t=20$ and $\eta=0.5$.  The white line shows the spatial extent of the active region, outside of which the MPS tensors correspond to those in Eq.~(\ref{eq:purification}).  The light-cone spreading of the entanglement is clear.  Note, this entanglement entropy is a property of the purified wavefunction, rather than of the physical density matrix.  Nonetheless, it illustrates the key numerical bottleneck in our calculation:  The peak entanglement entropy grows linearly in time, and hence the required bond dimension grows exponentially~\cite{kim2013}.
%and in principle there could be disentangling procedures which 

To calculate the current-current correlator, we relabel our sites to produce the translated state
$\hat T_x|\psi(0,\tau)\rangle=|\psi(x,\tau\rangle$.  The current-current correlator is then found by calculating the overlap between the shifted and original states,
\begin{equation}
\Lambda(x,\tau)=\langle\psi^*(0,\tau/2)|\psi(x,\tau/2)\rangle.
\end{equation}
In practice, the translation is implemented by temporarily appending un-evolved sites to the end of the chains. Given the product state on the wings of purified wavefunction, for $x>2l_0-1$ the correlator trivially factors into
\begin{equation}
\Lambda(x>2l_0-1,\tau)=\langle\psi^*(0,\tau/2)|\psi_0\rangle\langle \psi_0|\psi(x,\tau/2)\rangle,
\end{equation}
which vanishes due to the absence of equilibrium currents.  Figure~\ref{fig:ent}(b) shows the current-current correlator inside this expanding window.  One sees rapid oscillations, with timescale $\sim 1/U$, inside an envelope which decays on a timescale of order several $1/t$.  Visually, the current correlations do not appear to spread significantly, but instead remain confined to a central region of $\sim 20$ sites.

%In Fig.~\ref{fig:ent} we show heat maps of the spatially and temporally-resolved entanglement entropy (left) and current-current correlator (right) at $U/t=20$ and $\eta=0.5$. Solid lines show the size of the state over time: the white line bounding the entropy is equal to the number of bonds ($l-1$ for $l$ physical sites), while the black line is equal to the number of non-trivial translations ($2l-1$). Notably, while the current-current correlator vanishes quickly over time, the entanglement entropy increases monotonically in time and spreads out within an effective light cone. This growth of entanglement entropy, and the corresponding increase in bond dimension, is what constrains our simulations to moderate times.

\begin{figure}
    \centering
    \includegraphics[width=\columnwidth]{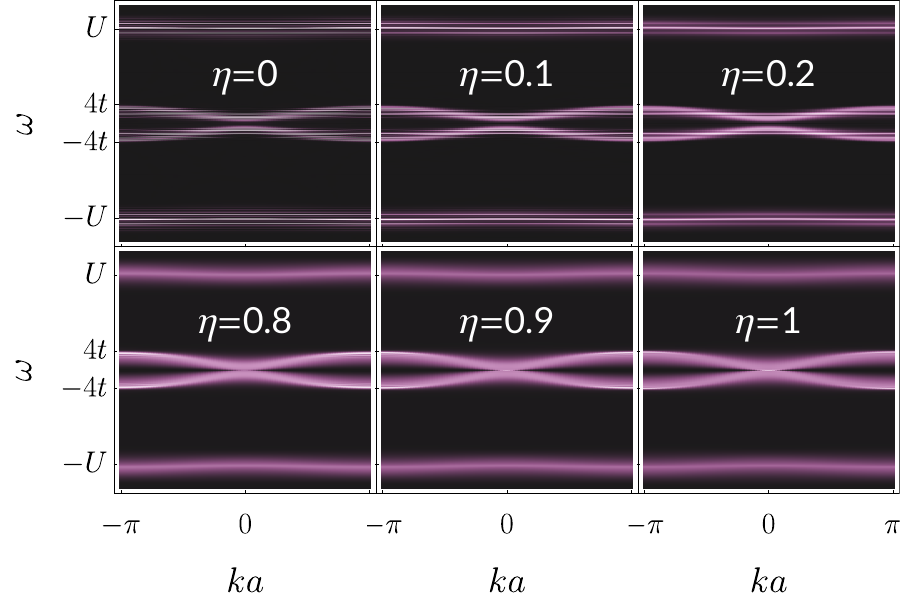}
    \caption{Infinite-temperature current-current correlation functions in the Fourier domain, $\Lambda(k,\omega)$, at $U/t=20$ and a range of imbalances. At $\eta=0$ we resolve a series of delta function contributions (see Sec.~\ref{sec:analytic}) that vary continuously with $k$. As argued in Appendix~\ref{sec:analyticDeets}, spectral weight is confined around $\omega=0$ and $\omega=\pm U$. As $\eta$ increases, these delta functions broaden. As $\eta\to 1$, sharp low-frequency features appear near $k=0$ and $k=\pm\pi$, while the $\omega\sim U$ excitations become weakly dispersive.}
    \label{fig:multipanel}
\end{figure}

By Fourier transforming the data in Fig.~\ref{fig:ent}, we can calculate $\Lambda(k,\omega)$, which gives the frequency and momentum dependence of the current correlations.  Figure~\ref{fig:multipanel} shows
%The power of this real-space approach is illustrated in Fig.~\ref{fig:multipanel}, where we plot 
$\Lambda(k,\omega)$ for $U/t=20$ and a variety of mass imbalances. One clearly sees structures on two frequency scales: low-frequency contributions bounded by $\pm 4t$, and high-frequency contributions near $\omega\sim \pm U$. 
This same structure is apparent in the time domain in Fig.~\ref{fig:ent}, and is described in more detail in Appendix~\ref{sec:analyticDeets} for $\eta=0$.
For small $\eta$ one sees a fine structure of bands, whose frequencies correspond to the delta-functions in the optical conductivity.    For $\eta=0$ the spectral weight associated with $\omega\sim U$ is independent of $k$.  This indicates that the associated excitations are localized.  These correspond to processes where the current operator has created or broken up a doublon.

At larger $\eta$ the momentum resolved spectra are smooth.  The bands with $\omega \sim \pm U$ are weakly dispersing, as the doublons can hop with an effective matrix element $t_{\rm eff}\sim \eta t^2/U$.  As $\eta\to 1$, we resolve a cusp in the optical conductivity near $\omega,k=0$ arising due to superdiffusive fluctuations at the highly-symmetric, integrable point~\cite{fava2020}.

According to Eq.~(\ref{eq:highTSigma}) the DC conductivity can be found from $\Lambda(k,\omega)$ by taking the limit $\omega\to0$ at fixed $k=0$, 
$\sigma_{DC}= \sigma_{\rm ref}\lim_{\omega\to 0} \Lambda(k=0,\omega)$, where $\sigma_{\rm ref}=a\beta t$. Conversely, at fixed $k\neq0$ one expects $\Lambda(k,\omega=0)=0$, as a  static longitudinal electric field with zero mean cannot generate equilibrium currents.  Thus the limits $k\to0$ and $\omega\to0$ do not commute.  This singular structure is smoothed out by our finite time effects.  Nonetheless, in the larger $\eta$ data in Fig.~\ref{fig:multipanel},  one sees different behavior when approaching the origin from different directions.  The apparent spectral gap at small but non-zero $\eta$ is a finite-time artifact.

%should be finite, with a value set by the Drude weight.  That is, if one first takes $k$ to zero, then $\omega$ to zero one finds the Drude weight.  Conversely $\lim_{k\to 0} \Lambda(k,\omega=0)=0$.  In the numerical data 

%From our finite-time expansion simulations in Fig.~\ref{fig:ent}, we resolve a sharp separation of frequency scales: low-frequency contributions bounded by $\pm 4t$, and high-frequency contributions near $\omega\sim U$. This separation of scales is derived analytically in Appendix~\ref{sec:analyticDeets} for $\eta=0$. Moreover, for $\eta=0$ we resolve the delta-function contributions to the optical conductivity (broadened by a finite time interval) whose arguments appear to vary continuously with momentum: $\delta(\omega\pm\omega_0(k))$. We additionally find that excitations with $\omega\sim U$ are ``flat", i.e. their contributions to the optical conductivity are momentum-independent. As $\eta$ is increased, the delta-functions broaden and merge while the $\omega\sim U$ branch becomes weakly dispersive. As $\eta\to 1$, we resolve a cusp in the optical conductivity near $\omega,k=0$ arising due to superdiffusive fluctuations at the highly-symmetric, integrable point~\cite{fava2020}.

\section{Power-Law Correlations\label{sec:kpz}}
Here we provide a derivation of the relationship between the dynamical scaling exponent, $z$, and the power-law exponent of the integrated current-current correlator, $\alpha$, presented in Sec.~\ref{sec:results}. Equivalent expressions can be found in \cite{bertini2021}.
%While these results are a standard application of hydrodynamic theory and appear elsewhere~\cite{bertini2021}, they are often quoted rather than derived.

Given that particle and spin densities are conserved in our system, a local density disturbance will spread out, with width $W^2(\tau)=\int dx~x^2\delta n_\uparrow(x,\tau)$ scaling as $W^2(\tau\to\infty)\propto\tau^{2/z}$, where $z$ is the dynamical exponent~\cite{dupont2020,bertini2021}.  For ballistic transport $z=1$, while for diffusive transport $z=2$.  This behavior is also found in the density-density correlation function,
\begin{equation}
    C_{\uparrow\uparrow}(x,\tau)=\frac{1}{N_s}\sum_i\langle n_\uparrow(i+x,\tau)n_\uparrow(i,0)\rangle,
\end{equation}
which by the fluctuation-dissipation theorem can be related to the density response of the system to a local potential.  As such, we can define
\begin{equation}\label{eq:sigmasquared}
    \Sigma_{\uparrow\uparrow}^2(\tau)=\sum_x x^2 C_{\uparrow\uparrow}(x,\tau)-\left(\sum_x x C_{\uparrow\uparrow}(x,\tau)\right)^2,
\end{equation}
which encodes the width of a density disturbance following a local perturbation.  Hence, it too obeys $\Sigma_{\uparrow\uparrow}^2(\tau)\propto\tau^{2/z}$ in the long time (hydrodynamic) limit~\cite{dupont2020,bertini2021}.

For notational simplity, we leave off the subscript $\uparrow$, and calculate the rate of change of $\partial_\tau^2\Sigma^2(\tau)=\sum_x x^2\partial_\tau^2 \langle n(x,\tau) n(x,0)\rangle$.  Using time-translational invariance and reflection symmetry, we can rewrite this as
\begin{equation}
\partial_\tau^2\Sigma^2(\tau)=
-\sum_x x^2 \langle \partial_{\tau_1}  n(x,\tau_1) \partial_{\tau_2}n(0,\tau_2)\rangle,
\end{equation}
evaluated at $\tau_1=\tau$ and $\tau_2=0$.  We then relate the time derivatives of the densities to the current, $\partial_\tau n(x,\tau)=j(x+1,\tau)-j(x,\tau)$, and rearrange the sum to connect $\Sigma^2$ to the current-current correlator:
\begin{equation}\label{eq:sigmalambda}
        \partial_\tau^2\Sigma^2(\tau)=
        2\sum_x \langle j(x,\tau)j(0,0)\rangle=2\Lambda(\tau).
\end{equation}
Given that $\Sigma^2(\tau)\propto\tau^{2/z}$, Eq.~(\ref{eq:sigmalambda}) yields $\Lambda(\tau)\propto\tau^{2/z-2}$. Hence, the integrated current-current correlator should scale as $\tau^\alpha$ with $\alpha=2/z-1$. This same exponent describes the associated low-frequency behavior of the optical conductivity: $\sigma(\omega)\propto|\omega|^{-\alpha}$~\cite{bertini2021}.  For generic mass imbalance we expect $z=2$ and hence $\alpha=0$.  At $\eta=1$, KPZ scaling predicts $z=3/2$ and hence $\alpha=1/3$.

\section{Details of Analytic Expansion\label{sec:analyticDeets}}

In the Falicov-Kimball limit $\eta\to 0$, the mass-imbalanced Hubbard Hamiltonian reduces to
\begin{multline}
    H_0=-t\sum_{i} \left(c^\dagger_{i,\uparrow}c_{i+1,\uparrow}+{\rm H.c.}\right)    +U\sum_in_{i,\uparrow}n_{i,\downarrow}.
\end{multline}
As noted in the main text, the Hamiltonian $H_0$ commutes with the $\downarrow$ particle density on every lattice site, $[H_0,n_{i,\downarrow}]=0$, so the $\downarrow$ spin densities on each site are conserved. 
%Given this, we can determine the current-current correlator for a fixed configuration of $\downarrow$ particles and then perform a disorder average to find the infinite-temperature properties. 
For a given configuration of $\downarrow$ spins, 
$\{n\}=\{n_{i,\downarrow}\}$,
%$\langle n_{i,\downarrow}\rangle$, 
the effective Hamiltonian for the $\uparrow$ spins is 
\begin{multline}
    \mathcal{H}_{0}=-t\sum_{i} \left(c^\dagger_{i,\uparrow}c_{i+1,\uparrow}+{\rm H.c.}\right)+\sum_iV(r_i)n_{i,\uparrow}
    \label{eq:h0_quadratic}
\end{multline}
where $V(r_i)=U n_{i,\downarrow}$. The fact that Eq.~(\ref{eq:h0_quadratic}) is quadratic in the fermion operators means that its properties can be exactly calculated.
%can be diagonalized for each configuration $\{n\}=\{n_{i,\downarrow}\}$. 
%This approach, however, becomes intractable when one must average over all possible configurations. Instead, we note that the four-point correlator $\Lambda(x,\tau)$ can generically be decomposed into the product of two-point correlators via Wick's theorem. Using the single-particle Green's functions defined in the main text, one can write
Using Wick's theorem, we write the current-current correlation function as
% \begin{multline}
% \Lambda_{\{n\}}(x,\tau)=\\-\frac{t^2}{N_s} \sum_{i,s_1 s_2}
% s_1 s_2\, G^>_{i,i+x}(\tau)
% G^<_{i+x+s_1,i+s_2}(-\tau)\label{ccgg}
% \end{multline}
% where the $s_j=\pm 1$ and $i$ is summed over all $N_s$ lattice sites.
\begin{equation}
\Lambda_{\{n\}}(x,\tau)=-\frac{t^2}{N_s} \sum_{\substack{i,j,k,l\\j-i=x}}
\eta_{ik}\eta_{jl}\, G^>_{i,j}(\tau)
G^<_{k,l}(-\tau)\label{ccgg}
\end{equation}
where
\begin{equation}
    \eta_{ik}=
    \begin{cases}
        1 & {\rm if}~k-i=1\\
        -1 & {\rm if}~k-i=-1\\
        0 & {\rm otherwise}
    \end{cases}
\end{equation}
and $i,j,k,l$ are summed over all $N_s$ lattice sites with the constraint that $j-i=x$.
Here $G^>_{ij}(\tau)=\langle c_i(\tau) c_j^\dagger(0)\rangle$ and $G^<_{ij}(\tau)=-\langle c_j^\dagger(0) c_i(\tau)\rangle$ are the single particle Greens functions, whose dependence on $\{n\}$ has been suppressed.
% We also introduce the Fourier transform
% \begin{equation}
% \Lambda_{\{n\}}(q,\tau)=\sum_x e^{-i q x} \Lambda_{\{n\}}(x,\tau).
% \end{equation}

Throughout we will consider the strong coupling limit where $U\gg t$.

\subsection{Decomposition into Regions of fixed $n_\downarrow$}

%As written, t
The configuration of $\downarrow$ spins acts as a binary disorder potential. %At infinite temperature, 
Even infinitesimal disorder should Anderson-localize all the single-particle wavefunctions. %For small $U/t$, however, the localization length will be quite large. We instead consider the limit of
In the regime of interest,
$U\gg t$, the localization length is very short, and the single-particle wavefunctions are confined
%completely %localized 
to regions of constant background potential -- ie. contiguous sequences of sites for which $n_\downarrow$ is constant.
% . For a given configuration $\{n_{i,\downarrow}\}$, then, 
Hence, the Green's functions should vanish unless $i$ and $j$ are in the same region, $\Omega$, and $k,l$ are in the same region, $\Omega^\prime$.  Due to the constraints from the $\eta$'s, we then have two possibilities:  either $\Omega=\Omega^\prime$, or $\Omega,\Omega^\prime$ are neighboring regions.  This leads to the decomposition %Given this constraint, 
%it is natural to 
%We
%rewrite Eq.~(\ref{ccgg}) by summing over the contributions from distinct regions, $\Omega$, of constant background potential,
%within a length-$N_s$ system. Formally, this replaces the dependence of $\Lambda_{\{n\}}$ on $\{n\}$. This yields
\begin{widetext}
\begin{equation}\label{eq:regiondecomp}
    \Lambda_{\{n\}}(x,\tau)=-\frac{t^2}{N_s} \sum_\Omega \sum_{\substack{i,j,k,l\in\Omega\\
    j-i=x}} \eta_{ik}\eta_{jl}\, G^>_{i,j}(\tau)
    G^<_{k,l}(-\tau)
    -\frac{t^2}{N_s}\sum_{\langle\Omega,\Omega'\rangle}\sum_{\substack{i,j\in\Omega\\k,l\in\Omega'\\j-i=x}}
    \eta_{ik}\eta_{jl}\, \left(G^>_{i,j}(\tau)
    G^<_{k,l}(-\tau)
    +G^>_{k,l}(\tau)
    G^<_{i,j}(-\tau)
    \right)
\end{equation}
%We emphasize that this decomposition of Eq.(\ref{ccgg}) is {\it exact} given the aforementioned constraint on $G_{ij}$. 
%The terms within each sum
%, by construction, 
%depend 
%only on the size of the regions $\Omega$. %Defining these lengths as $m={\rm dim}(\Omega)$ and $m'={\rm dim}(\Omega')$, we 
%\begin{widetext}

which can be expressed as
\begin{equation}\label{eq:lambdadefns}
    \Lambda_{\{n\}}(x,\tau)=\frac{1}{N_s} \sum_\Omega m\Lambda_m^{(0)}(x,\tau)
    +\frac{1}{N_s}\sum_{\langle \Omega,\Omega'\rangle}\left(\frac{m+m'}{2}\right)\Lambda_{mm'}^{(U)}(x,\tau),
\end{equation}
% where $\langle\Omega,\Omega^\prime\rangle$ denotes a sum over nearest-neighbor regions:  The $\eta$ factors in Eq.~(\ref{eq:regiondecomp}) vanish for all other combinations.  
where $m$ is the length of segment $\Omega$ and $m'$ is the length of segment $\Omega'$. 
The summands only depend on the lengths of the segments,
%\begin{widetext}
\begin{eqnarray}\label{l0}
\Lambda_m^{(0)}(x,\tau)&=&
-\frac{t^2}{m}\sum_{\substack{i,j,k,l=1\\j-i=x}}^m
\eta_{ik}\eta_{jl}G^>_{ij}(\tau)G^<_{kl}(-\tau)\\\label{l2}
\Lambda^{(U)}_{mm^\prime}(x,\tau)
&=& -\frac{2t^2}{m+m^\prime}
%\sum_{i,j=1}^m \sum_{k,l=1}^{m^\prime}
\delta_{x,0}\left(
G^>_{11}(\tau)G^<_{00}(-\tau)+G^>_{00}(\tau)G^<_{11}(-\tau)\right).
\label{lu}
\end{eqnarray}
In Eq.~(\ref{l0}) we have used translational invariance to take the region $\Omega=[1,2,\cdots,m]$ to extend between sites $1$ through $m$.  In Eq.~(\ref{l2}) we take $\Omega$ to be to the left of $\Omega^\prime$, with $\Omega=[-m+1,-m+2,\ldots,0]$ and $\Omega^\prime=[1,2,\ldots,m']$. %respectively of length $m$ and $m'$.  
The factors of $m$ and $m'$ in Eq.~(\ref{eq:lambdadefns}) are chosen for notational convenience: $\sum_\Omega m=N_s$ and $\sum_{\langle \Omega,\Omega'\rangle}(m+m') = 2 N_s
$, as long as $\{n\}$ contains at least 2 regions.  

We now wish to rewrite the sum of $\Omega$ in Eq.~(\ref{eq:lambdadefns}) as a sum over $m$.  We note that
at infinite temperature, each site is independent and the probability of any given site containing a heavy ($\downarrow$) particle is equal to $\bar n_\downarrow$, the average density of heavy particles.  Consequently, the probability that a given site is in region containing $m$ adjacent $\downarrow$ spins is $P_m=m (1-\bar n_\downarrow)^2 \bar n_\downarrow^m$.  The factor of $m$ accounts for the $m$ different possible ways the region could contain that site.  The $ \bar n_\downarrow^m$ is due to the fact that we need $m$ sites which each contain a $\downarrow$ spin.  The $(1-\bar n_\downarrow)^2$ comes from the fact that the region must be bookended by empty sites.  Similarly the probability that a site is in a region of $m$ empty sites is $\bar P_m=m \bar n_\downarrow^2 (1-\bar n_\downarrow)^m$.  One can readily verify that $\sum_m (P_m+\bar P_m)=1$.  Finally we note that the total number of regions of length $m$ should be $\mathcal{N}_m= N_s (P_m+\bar P_m)/m$ and for any function, $f_m$,
% , and the fraction of regions which are of length $m$ is 
% \begin{equation}
% \rho_m= m \left[(1-\bar n_\downarrow)^2 \bar n_\downarrow^m+
% n_\downarrow^2 (1-\bar n_\downarrow)^m
% \right],
% \end{equation}
% where $\sum_m\rho_m=1$.  Thus 
\begin{equation}
\sum_\Omega f_m=\sum_m \mathcal{N}_m f_m=N_s \sum_m (P_m+\bar P_m) \frac{f_m}{m}
\end{equation}

We can apply an analogous argument to thinking about consecutive regions with $\Omega$ of length $m$ to the left of $\Omega^\prime$ of length $m^\prime$. If $\Omega$ is comprised of $\downarrow$ spins and $\Omega'$ of empty sites, the probability that an individual site is within this configuration is $P_{mm^\prime}=(m+m^\prime)\bar n_\downarrow^{m+1}(1-\bar n_\downarrow)^{m^\prime+1}$.
% An individual site is within such a configuration with probability $P_{mm^\prime}=(m+m^\prime)\left(\bar n_\downarrow^{m+1}(1-\bar n_\downarrow)^{m^\prime+1} + \bar n_\downarrow^{m+1}(1-\bar n_\downarrow)^{m^\prime+1}\right)$.
The probability of a site being in the complementary configuration (where $\Omega$ is comprised of empty sites and $\Omega'$ of $\downarrow$ spins) is $\bar P_{mm^\prime}=(m+m^\prime)\bar n_\downarrow^{m'+1}(1-\bar n_\downarrow)^{m+1}$.
Hence the total number of such configurations is ${\cal N}_{mm^\prime}= N_s (P_{mm^\prime}+\bar P_{mm^\prime})/(m+m^\prime)$, and
\begin{equation}
\sum_{\langle \Omega, \Omega^\prime\rangle} f_{mm^\prime}=\sum_{mm^\prime} {\cal N}_{mm^\prime} f_{mm^\prime}=N_s \sum_{mm^\prime} \left(P_{mm^\prime}+{\bar P}_{mm^\prime}\right)\frac{f_{mm^\prime}}{m+m^\prime}.
\end{equation}
% , and the fraction of consecutive regions with this property are
% \begin{equation}
% \rho_{mm^\prime}=
% \left(\frac{m+m'}{2}\right)\left(\bar n_\downarrow^{m+1}(1-\bar n_\downarrow)^{m'+1}+(1-\bar n_\downarrow)^{m+1}\bar n_\downarrow^{m'+1}\right)\label{eq:rhommp}
% \end{equation}
% with $\sum_{mm^\prime}\rho_{mm^\prime}=1$. The factor of $1/2$ in Eq.~(\ref{eq:rhommp}) comes from the fact that each region $\Omega$ is double-counted in the sum over adjacent regions $\langle\Omega,\Omega'\rangle$.

We can then rewrite Eq.~(\ref{eq:lambdadefns}) as
\begin{equation}\label{eq:lambda_rho_final}
\Lambda(x,\tau)=\underbrace{\sum_m (P_m+\bar P_m)\Lambda^{(0)}_m(x,\tau)}_{\text{\normalsize $\Lambda^{(0)}(x,\tau)$}}+\underbrace{\frac{1}{2}\sum_{m,m'}\left(P_{mm^\prime}+\bar P_{mm^\prime}\right)\Lambda^{(U)}_{mm'}(x,\tau)}_{\text{\normalsize $\Lambda^{(U)}(x,\tau)$}}
\end{equation}
%where we define $\rho_m=\mathcal{N}_m/N_s$ and $\rho_{mm'}=\mathcal{N}_{mm'}/N_s$. 
In the subsequent sections, we will evaluate the terms $\Lambda^{(0)}_m$ and $\Lambda^{(U)}_{mm'}$ analytically in the limit $U\gg t$. This calculation is more natural in momentum space, so we introduce the Fourier transform $\Lambda(q,\tau)=\sum_xe^{iqx}\Lambda(x,\tau)$.

\subsection{Spectral Representation}\label{spec}
%We will now define a spectral representation for the Green's functions. 
At infinite temperature, $G_{ij}^>(\tau)=(1-\bar n_\uparrow) A_{ij}(\tau)$ and $G_{ij}^<(\tau)=-\bar n_\uparrow A_{ij}(\tau)$, with spectral function $A_{ij}(\tau)=\sum_\nu \psi_i^\nu (\psi_j^\nu)^* e^{-i\epsilon_\nu \tau}$. Here we use $\nu$ to label the single particle eigenstates. In our case the $\uparrow$ particle density is $\bar n_\uparrow=1/2$.

Let's now consider a region of $m$ sites 
% in a region 
with $V=0$, surrounded by sites with $V=U$. We'll treat these as effective hard-wall boundary conditions, labeling the sites in the constant-potential region as $i=1,2,\ldots, m$. The single-particle eigenstates are given by $\psi_i^n= A \sin{ki}$ where $k=\pi n/(m+1)$; the integer $n$ can take values $1,2,\ldots,m$. Note that we will mix our notation and use $k$ and $n$ interchangeably, for example, defining the energy $\nu_n=-2t\cos{k}$, which should be interpreted as $\nu_n=-2t \cos \pi n/(m+1)$.
%making reference to these definitions. The energy is $\nu_n=-2t\cos{k}$, and t
The normalization factor is $|A|^2=2/(m+1)$. Eigenstates in an analogous high-potential region are identical, with energies just shifted by $U$.

One consequence is that $\Lambda^{(0)}$ captures the dominant low-frequency contributions to $\Lambda$, while $\Lambda^{(U)}$ contains frequency components which are of order $U$.

% The low frequency contribution, $\Lambda^{(0)}$ comes from terms in Eq.~(\ref{ccgg}) where $i,i+x,i+x+s_1$, and $i+s_2$ are all in the same region.  Conversely, the high frequency contribution, $\Lambda^{(U)}$, involves terms where $i,i+x$ are in one region, but $i+x+s_1,i+s_2$ are in another.  In the following two sections we separately calculate each of these, and then $\Lambda=\Lambda^{(0)}+\Lambda^{(U)}$.

% {\color{red} Write $\Lambda^{(0)}$ in terms of each region}

%We will write $\Lambda=\Lambda^{(0)}+\Lambda^{(U)}$, where the $\Lambda^{(0)}$ oscillates with frequencies that are small compared to $U$.

\subsection{Low-Frequency Contributions: $\Lambda^{(0)}$}
We first compute the low-frequency contributions $\Lambda_m^{(0)}(q,\tau)=\sum_x e^{iqx}\Lambda_m^{(0)}(x,\tau)$ from regions of size $m$ to the current-current correlator (see Eq.~(\ref{eq:lambda_rho_final})). To begin, we return to the definition of $\Lambda^{(0)}_m$ in Eq.~(\ref{l0}).
% s.~(\ref{eq:regiondecomp}) and~(\ref{eq:lambdadefns}).
%We will now compute the contribution to $\Lambda^{(0)}(q,\tau)$, 
% As already explained,
% the low-frequency contribution, $\Lambda^{(0)}(q,\tau)$, 
% corresponds to the sum of terms in 
% %by requiring that all Green's functions in 
% Eq.~(\ref{slam}) where all of the Green's function indices are in the same region, which we will take to have size $m$.
% %We wi have their indices within the aforementioned region of size $m$. 
% It is therefore natural to define
% \begin{widetext}
% \begin{equation}
% \Lambda_m^{(0)}(q,\tau)=-\frac{t^2}{m}\sum_{i,s_1,s_2} e^{iqx}G^>_{i,i+x}(\tau)G^<_{i+x+s_1,i+s_2}(-\tau)
% \end{equation}
% \begin{equation}
% \Lambda_m^{(0)}(q,\tau)=-\frac{t^2}{m}\sum_{l,j=1}^{m-1} e^{iq(l-j)} \left[
% G^>_{l,j}(\tau)G^<_{j+1,l+1}(-\tau)
% -G^>_{l+1,j}(\tau) G^<_{j+1,l}(-\tau)
% +G^>_{l+1,j+1}(\tau) G^<_{j,l}(-\tau)
% -G^>_{l,j+1}(\tau)G^<_{j,l+1}(-\tau)
% \right].
% \end{equation}
% Inserting explicit factors of the density $\bar n_\uparrow=1/2$ 
% and e
Evaluating the factors of $\eta_{ik}$,
%this contribution is given by
% \begin{multline}
% \begin{equation}
% \begin{split}
% \Lambda^{(0)}_m(q,\tau)=\frac{-t^2}{4}\sum_{l,j=1}^{m-1} e^{iq(l-j)}&
% \bigg(A_{l,j}(\tau) A_{j+1,l+1}(-\tau)\\-&A_{l+1,j}(\tau)A_{j+1,l}(-\tau)\\ + &A_{l+1,j+1}(\tau) A_{j,l}(-\tau)\\ - &A_{l,j+1}(\tau)A_{j,l+1}(-\tau) \bigg).
% % \end{multline}
% \end{split}
%\end{equation}
% \begin{widetext}
\begin{equation}
%\begin{split}
\Lambda^{(0)}_m(q,\tau)=\frac{t^2\bar n_\uparrow(1-\bar n_\uparrow)}{m}\sum_{l,j=1}^{m-1} e^{iq(l-j)}
\bigg(A_{l,j}^+ A_{j+1,l+1}^-
-A_{l+1,j}^+A_{j+1,l}^- + A_{l+1,j+1}^+ A_{j,l}^- - A_{l,j+1}^+A_{j,l+1}^- \bigg).
% \end{multline}
%\end{split}
\end{equation}
where $A_{l,j}^+=A_{l,j}(\tau)$
and $A_{l,j}^-=A_{l,j}(-\tau)$.
If we expand $A_{l,j}$ in terms of single-particle wavefunctions and collect terms with the same site index, this can be rewritten as
% \begin{multline}
% \begin{equation}
% \begin{split}
%  \Lambda^{(0)}_m(q,\tau)= \frac{-t^2}{4}\sum_{n\bar n} e^{i (\nu_n-\nu_{\bar n})\tau}\bigg(&C_{\bar n n}(q) C_{\bar n n}(-q)\\ + &C_{n \bar n}(q) C_{n \bar n}(-q)\\ - &C_{n \bar n}(q) C_{\bar n n}(-q) \\- &C_{\bar n n}(q) C_{n \bar n}(-q)\bigg).
%  \label{dif}
% % \end{multline}
% \end{split}
% \end{equation}
\begin{equation}
 \Lambda^{(0)}_m(q,\tau)= \frac{t^2\bar n_\uparrow(1-\bar n_\uparrow)}{m}\sum_{n,\bar n=1}^m e^{i (\nu_n-\nu_{\bar n})\tau}\bigg(C_{\bar n n}(q) C_{\bar n n}(-q) + C_{n \bar n}(q) C_{n \bar n}(-q) - C_{n \bar n}(q) C_{\bar n n}(-q) - C_{\bar n n}(q) C_{n \bar n}(-q)\bigg).
 \label{dif}
% \end{multline}
\end{equation}
where $C_{\bar n n}(q)= \sum_{l=0}^m e^{ilq} \psi_l^{\bar n} \psi_{l+1}^{n}$. Evaluating $\Lambda^{(0)}_m(q,\tau)$ now amounts to evaluating $C_{\bar n n}(q)$. Inserting the explicit form of the single-particle wavefunctions and summing over $l$ yields
% \begin{multline}
% \begin{equation}
% \begin{split}
%  C_{\bar n n}(q)
%  = -\frac{A^2}{4} 
%  \bigg[
% &\frac{1-e^{i (k+\bar k+q) (m+1)}}{1-e^{i (k+\bar k+q)}}e^{i \bar k} \\
% &+
% \frac{1-e^{-i (k+\bar k-q) (m+1)}}{1-e^{-i (k+\bar k-q)}}e^{-i \bar k} \\
% &-\frac{1-e^{i (k-\bar k+q) (m+1)}}{1-e^{i (k-\bar k+q)}}e^{-i \bar k} \\
% &-
% \frac{1-e^{-i (k-\bar k-q) (m+1)}}{1-e^{-i (k-\bar k-q)}}e^{i \bar k} 
%  \bigg].
%  \label{eq:cnnsum}
% % \end{multline}
% \end{split}
% \end{equation}
\begin{equation}
 C_{\bar n n}(q)
 = -\frac{A^2}{4} 
 \bigg[
 \Upsilon(k+\bar k+q)
%\frac{1-e^{i (k+\bar k+q) (m+1)}}{1-e^{i (k+\bar k+q)}}
e^{i \bar k} 
+
\Upsilon(-k-\bar k+q)
%\frac{1-e^{-i (k+\bar k-q) (m+1)}}{1-e^{-i (k+\bar k-q)}}
e^{-i \bar k} -
\Upsilon(k-\bar k+q)
%\frac{1-e^{i (k-\bar k+q) (m+1)}}{1-e^{i (k-\bar k+q)}}
e^{-i \bar k} -
\Upsilon(-k+\bar k+q)
%\frac{1-e^{-i (k-\bar k-q) (m+1)}}{1-e^{-i (k-\bar k-q)}}
e^{i \bar k} 
 \bigg].
 \label{eq:cnnsum}
% \end{multline}
\end{equation}
%\end{widetext}
with
\begin{equation}
\Upsilon(p)=\frac{1-e^{i  (m+1)p}}{1-e^{i p}}.
\end{equation}
Following the notation introduced in Sec.~\ref{spec}, $k=\pi n/(m+1)$ and $\bar k=\pi \bar n/(m+1)$.
While the expression in Eq.~(\ref{eq:cnnsum}) can be used to numerically calculate the response function for arbitrary $q$, it is cumbersome to work with analytically.
We can, however, simplify it when $q=0$ or $q=\pi$. The former results are quoted in the main text.

We first take $q=0$.
Note that because of the form of the momentum the phase factor in the numerator obeys $e^{i(m+1)p}=\pm 1$.  If $p=0$, corresponding to $n=\bar n$, then both the numerator and denominator vanish, and by using L'H\^opital's rule, we see that $\Upsilon(0)=m+1$.  Otherwise $\Upsilon(p)$ is non-zero only for odd $p$, corresponding to $n$ and $\bar n$ having opposite parity.  The $p=0$ terms cancel with one-another when substituted into Eq.~(\ref{dif}), so we only need to consider the terms where $n$ and $\bar n$ have opposite parity, ie. $e^{\pm i (k\pm\bar k) (m+1)}=-1$.  After a bit of algebra, one finds
%If we take $q=0$, then $\Upsilon(p)=0$ unless $e^{i(m+1)p}$
%we can restrict $n,\bar n$ to have opposite parity. This is because, when $n$ and $\bar n$ have the same parity, $e^{\pm i (k\pm\bar k) (m+1)}=1$. This includes the apparently singular case of $\bar n = n$, which clearly does not contribute to Eq.~(\ref{dif}). As discussed in the main text, this parity constraint  results in the depletion of spectral weight near $\omega=0$. When $n$ and $\bar n$ have differing parity, $e^{\pm i (k\pm\bar k) (m+1)}=-1$ and, with a bit of algebra, one finds
\begin{equation}
C_{\bar n n}(0)
=\frac{1}{m+1}\left(
\frac{\sin (k-\bar k)/2}{\sin(k+\bar k)/2}-
\frac{\sin (k+\bar k)/2}{\sin(k-\bar k)/2}\right).\qquad \mbox{$(n+\bar n)$ odd}
\end{equation}
Clearly $C_{\bar n n}(0)$ is odd when one switches $n$ and $\bar n$, so the contribution to the uniform current-current correlator can be written as
% \begin{equation}
% \Lambda^{(0)}_m(0,\tau)= -t^2 \sum_{n\bar n}^\prime e^{i (\nu_n-\nu_{\bar n})\tau} C_{\bar n n}^2. \label{eq:lambda0m_q0}
% \end{equation}
\begin{equation}
\Lambda^{(0)}_m(0,\tau)= \frac{4t^2\bar n_\uparrow(1-\bar n_\uparrow)}{m} \sum_{n\bar n}^\prime e^{i (\nu_n-\nu_{\bar n})\tau} \left(C_{\bar n n}(0)\right)^2. \label{eq:lambda0m_q0}
\end{equation}
where the prime indicates that we just include terms with opposite parity.
The full low-energy contribution can now be found by inserting Eq.~(\ref{eq:lambda0m_q0}) into Eq.~(\ref{eq:lambda_rho_final}). To calculate the correlator numerically, as we do in the main text, we simply truncate the $m$ sum to be finite. In Fig.~\ref{fig:timeseriesFig}, we truncate $\Lambda^{(0)}$ at $m=30$.

% The full low-energy contribution can now be found by inserting Eq.~(\ref{eq:lambda0m_q0}) into Eq.~(\ref{eq:lambda_rho_final}):
% \begin{equation}\label{combine}
% \Lambda^{(0)}(q,\tau) = \sum_m \rho_m\Lambda_m^{(0)}(q,\tau).
% \end{equation}
% where $\rho_m$ is defined in Eq.~(\ref{eq:rhom}).
% The contribution to $\Lambda^{(0)}$ per site is $\Lambda_m^{(0)}/m$.  Thus the total $\Lambda^{(0)}$ per site is
% \begin{equation}\label{combine}
% \Lambda^{(0)}(0,\tau) = \sum_m p_m \frac{\Lambda_m^{(0)}(0,\tau)}{m}
% \end{equation}
% where $p_m$ is the probability that a given site is in a segment of length $m$.  In the main text, the expression is given for arbitrary density of heavy particles, $\bar n_\downarrow$, although we limit ourselves to $\bar n_\downarrow=1/2$.  The probability of the site in question to be in a run of length $m$ of heavy particles is $m n_\downarrow^m (1-n_\downarrow)^2$:  There are $m$ such runs, each of which requires $m$ heavy particles, flanked by 2 empty sites.  The probability of being in an empty run of length $m$ is $m (1-n_\downarrow)^m n_\downarrow^2$. Thus
% \begin{eqnarray}
% p_m &=& m (n_\downarrow^m (1-n_\downarrow)^2+(1-n_\downarrow)^m n_\downarrow^2)
% \end{eqnarray}
% To calculate the correlator numerically, as we do in the main text, we simply truncate the $m$ sum to be finite.

A similar argument holds for $\Lambda_m^{(0)}(\pi,\tau)$, and we find
\begin{equation}
    \Lambda_m^{(0)}(\pi,\tau)=\frac{t^2\bar n_\uparrow(1-\bar n_\uparrow)}{m}\sum_{n=1}^m e^{-2i\tau\nu_n}\nu_n^2.
\end{equation}
%{\color{blue}The full correlator can be constructed from these $m$-dependent contributions using Eq.~(\ref{eq:lambda_rho_final}).}  
Note, that this only involves frequencies which are an even multiple of $\pi/(m+1)$.  Conversely Eq.~(\ref{eq:lambda0m_q0}) only involves odd multiples.

\subsection{High-Frequency Contributions: $\Lambda^{(U)}$}
Here we compute the high-frequency contributions $\Lambda_{m,m'}^{(U)}(q,\tau)=\sum_x e^{iqx}\Lambda_{m,m'}^{(U)}(x,\tau)$ from adjacent regions of size $m,m'$ to the current-current correlator. 
Due to the Kronecker delta on in Eq.~(\ref{lu}), $\Lambda_{m,m'}^{(U)}(q,\tau)=\Lambda^{(U)}_{m,m'}(\tau)$ is independent of $q$.  This structure is seen in Fig.~\ref{fig:multipanel}.  

%Contributions to this correlator come only from the boundaries between adjacent regions. Hence, at this level in perturbation theory, the spectral weight around $\omega\sim U$ is non-dispersive (independent of momentum $q$). In what follows, we omit the $q$-dependence for notational simplicity: $\Lambda^{(U)}_{m,m'}(q,\tau)\equiv\Lambda^{(U)}_{m,m'}(\tau)$. 

% To begin, we return to the definition of $\Lambda^{(0)}_m$ in Eqs.~(\ref{eq:regiondecomp}) and~(\ref{eq:lambdadefns}).

% The high frequency structure comes from configurations where $i$ and $j$ are in different segments, and $i+s_2$ and $j+s_1$ are in the segments of $j$ and $i$. Thus, at this level in perturbation theory, the spectral weight around $\omega\sim\pm U$ is non-dispersive (independent of momentum $q$). 

As already introduced, we take the left segment, $\Omega$, to be of length $m$ and the right segment to be of length $m^\prime$. We will label the left sites as $j=-m+1,-m+2,\cdots,0$ and the right one as $i=1,2,\cdots,m^\prime$.  
Since $\Lambda^{(U)}_{mm^\prime}=\Lambda^{(U)}_{m^\prime m}$
there is no loss of generality in taking the right segment to be the one with heavy particles.  
The energy eigenstates in the left and right regions are $\phi_j^n$ and $\psi_i^{n^\prime}$, 
%and $\phi_j^n$
%with $j=-m+1,\ldots, 0$ and $i=1,2,\ldots,m$,
\begin{equation}
\phi_j^n=\sqrt{\frac{2}{m+1}}\sin k (1-j),\quad
\psi_i^{n^\prime}=\sqrt{\frac{2}{m^\prime+1}}\sin k^\prime i.
\end{equation}
Here
%for site indices $i=1,2,\ldots,m$ and
$k=\pi n/(m+1),k^\prime=\pi {n^\prime}/(m^\prime+1)$, and the eigen-energies are $\nu_n=-2t\cos(k),\mu_{n^\prime}=U-2t\cos(k^\prime)$.  
%The left wavefunctions are
%\begin{equation}
%\phi_j^n=\sqrt{\frac{2}{m+1}}\sin k (1-j)
%\end{equation}
%for site indices $j=-m+1,\ldots, 0$ with $k=\pi n/(m+1)$.  Their energies are $\nu_n=-2t\cos(k)$. 
The correlator only depends on the values of the wavefunctions at the boundary between the regions:
\begin{equation}
\phi_0^{n}=\sqrt{\frac{2}{m+1}}\sin k,\quad\psi_1^{n^\prime}=\sqrt{\frac{2}{m^\prime+1}}\sin k^\prime. 
\end{equation}
%The contribution to the current-current correlator is
% \begin{widetext}
In particular,
\begin{eqnarray}
\Lambda_{m,m^\prime}^{(U)}(\tau)&=& -\frac{2t^2}{m+m'} (G^>_{11}(\tau)G^<_{0,0}(-\tau) + G^>_{00}(\tau)G^<_{1,1}(-\tau))\\
&=&\frac{2t^2\bar n_\uparrow (1-\bar n_\uparrow)}{m+m'} \sum_{n=1}^m\sum_{n^\prime=1}^{m^\prime}
\left(e^{i(\mu_{n^\prime}-\nu_n)\tau}
+ e^{-i(\mu_{n^\prime}-\nu_n)\tau}\right)
|\psi^{n^\prime}_1|^2 |\phi_0^n|^2.
\end{eqnarray}
The full high-frequency correlator, $\Lambda^{(U)}(\tau)$, can be found by 
inserting this result into Eq.~(\ref{eq:lambda_rho_final}). In Fig.~\ref{fig:timeseriesFig}, we truncate $\Lambda^{(U)}$ at $m=m'=20$.

% summing over $m,m'$, weighting with the density determined in Eq.~(\ref{eq:rhommp}):
% \begin{equation}
% \Lambda^{(U)}(\tau) = \sum_{m,m'} \rho_{m,m'}\Lambda_{m,m'}^{(U)}(\tau).
% \end{equation}
\end{widetext}

% As a useful check, we can verify that our expressions satisfy the f-sum rule. For $\bar n_\uparrow=\bar n_\downarrow=\bar n/2$, this states that $\Lambda(\tau=0)=t^2 \bar n(2-\bar n)/2$. With a little algebra, we find indeed that
% \begin{equation}
% \begin{split} 
% \Lambda(\tau=0)&=\Lambda_0(\tau=0)+\Lambda_U(\tau=0) \\
% &=\frac{t^2}{4}\bar n(2-\bar n)(2-\bar n(2-\bar n))+\frac{t^2}{4}\bar n^2(2-\bar n)^2\\
% &=t^2 \bar n(2-\bar n)/2.
% \end{split}
% \end{equation}

\bibliographystyle{apsrev4-2}
\bibliography{bib}

\end{document}